\documentclass[ALICE,manyauthors]{cernphprep}
\usepackage[comma,square,numbers,sort&compress]{natbib}
\usepackage{hyperref}
\usepackage{lineno}
\usepackage{xspace}
\usepackage{amsmath}
\usepackage{amsfonts}
\usepackage{amssymb}
\usepackage{rotating}
\usepackage{multirow}
\usepackage{dcolumn}
\newcolumntype{d}[1]{D{.}{.}{#1}}
\usepackage[T1]{fontenc}
\usepackage{orcidlink} 

\begin{document}
%

\newcommand{\pp}           {pp\xspace}
\newcommand{\ppbar}        {\mbox{$\mathrm {p\overline{p}}$}\xspace}
\newcommand{\XeXe}         {\mbox{Xe--Xe}\xspace}
\newcommand{\PbPb}         {\mbox{Pb--Pb}\xspace}
\newcommand{\pA}           {\mbox{pA}\xspace}
\newcommand{\pPb}          {\mbox{p--Pb}\xspace}
\newcommand{\AuAu}         {\mbox{Au--Au}\xspace}
\newcommand{\dAu}          {\mbox{d--Au}\xspace}

\newcommand{\s}            {\ensuremath{\sqrt{s}}\xspace}
\newcommand{\snn}          {\ensuremath{\sqrt{s_{\mathrm{NN}}}}\xspace}
\newcommand{\pt}           {\ensuremath{p_{\rm T}}\xspace}
\newcommand{\meanpt}       {$\langle p_{\mathrm{T}}\rangle$\xspace}
\newcommand{\ycms}         {\ensuremath{y_{\rm CMS}}\xspace}
\newcommand{\ylab}         {\ensuremath{y_{\rm lab}}\xspace}
\newcommand{\etarange}[1]  {\mbox{$\left | \eta \right |~<~#1$}}
\newcommand{\yrange}[1]    {\mbox{$\left | y \right |~<~#1$}}
\newcommand{\dndy}         {\ensuremath{\mathrm{d}N_\mathrm{ch}/\mathrm{d}y}\xspace}
\newcommand{\dndeta}       {\ensuremath{\mathrm{d}N_\mathrm{ch}/\mathrm{d}\eta}\xspace}
\newcommand{\avdndeta}     {\ensuremath{\langle\dndeta\rangle}\xspace}
\newcommand{\dNdy}         {\ensuremath{\mathrm{d}N_\mathrm{ch}/\mathrm{d}y}\xspace}
\newcommand{\Npart}        {\ensuremath{N_\mathrm{part}}\xspace}
\newcommand{\Ncoll}        {\ensuremath{N_\mathrm{coll}}\xspace}
\newcommand{\dEdx}         {\ensuremath{\textrm{d}E/\textrm{d}x}\xspace}
\newcommand{\RpPb}         {\ensuremath{R_{\rm pPb}}\xspace}

\newcommand{\nineH}        {$\sqrt{s}~=~0.9$~Te\kern-.1emV\xspace}
\newcommand{\seven}        {$\sqrt{s}~=~7$~Te\kern-.1emV\xspace}
\newcommand{\twoH}         {$\sqrt{s}~=~0.2$~Te\kern-.1emV\xspace}
\newcommand{\twosevensix}  {$\sqrt{s}~=~2.76$~Te\kern-.1emV\xspace}
\newcommand{\five}         {$\sqrt{s}~=~5.02$~Te\kern-.1emV\xspace}
\newcommand{\twosevensixnn}{$\sqrt{s_{\mathrm{NN}}}~=~2.76$~Te\kern-.1emV\xspace}
\newcommand{\fivenn}       {$\sqrt{s_{\mathrm{NN}}}~=~5.02$~Te\kern-.1emV\xspace}
\newcommand{\LT}           {L{\'e}vy-Tsallis\xspace}
\newcommand{\GeVc}         {Ge\kern-.1emV/$c$\xspace}
\newcommand{\MeVc}         {Me\kern-.1emV/$c$\xspace}
\newcommand{\TeV}          {Te\kern-.1emV\xspace}
\newcommand{\GeV}          {Ge\kern-.1emV\xspace}
\newcommand{\MeV}          {Me\kern-.1emV\xspace}
\newcommand{\GeVmass}      {Ge\kern-.2emV/$c^2$\xspace}
\newcommand{\MeVmass}      {Me\kern-.2emV/$c^2$\xspace}
\newcommand{\lumi}         {\ensuremath{\mathcal{L}}\xspace}

\newcommand{\ITS}          {\rm{ITS}\xspace}
\newcommand{\TOF}          {\rm{TOF}\xspace}
\newcommand{\ZDC}          {\rm{ZDC}\xspace}
\newcommand{\ZDCs}         {\rm{ZDCs}\xspace}
\newcommand{\ZNA}          {\rm{ZNA}\xspace}
\newcommand{\ZNC}          {\rm{ZNC}\xspace}
\newcommand{\SPD}          {\rm{SPD}\xspace}
\newcommand{\SDD}          {\rm{SDD}\xspace}
\newcommand{\SSD}          {\rm{SSD}\xspace}
\newcommand{\TPC}          {\rm{TPC}\xspace}
\newcommand{\TRD}          {\rm{TRD}\xspace}
\newcommand{\VZERO}        {\rm{V0}\xspace}
\newcommand{\VZEROA}       {\rm{V0A}\xspace}
\newcommand{\VZEROC}       {\rm{V0C}\xspace}
\newcommand{\Vdecay} 	   {\ensuremath{V^{0}}\xspace}

\newcommand{\ee}           {\ensuremath{e^{+}e^{-}}} 
\newcommand{\pip}          {\ensuremath{\pi^{+}}\xspace}
\newcommand{\pim}          {\ensuremath{\pi^{-}}\xspace}
\newcommand{\kap}          {\ensuremath{\rm{K}^{+}}\xspace}
\newcommand{\kam}          {\ensuremath{\rm{K}^{-}}\xspace}
\newcommand{\pbar}         {\ensuremath{\rm\overline{p}}\xspace}
\newcommand{\kzero}        {\ensuremath{{\rm K}^{0}_{\rm{S}}}\xspace}
\newcommand{\lmb}          {\ensuremath{\Lambda}\xspace}
\newcommand{\almb}         {\ensuremath{\overline{\Lambda}}\xspace}
\newcommand{\Om}           {\ensuremath{\Omega^-}\xspace}
\newcommand{\Mo}           {\ensuremath{\overline{\Omega}^+}\xspace}
\newcommand{\X}            {\ensuremath{\Xi^-}\xspace}
\newcommand{\Ix}           {\ensuremath{\overline{\Xi}^+}\xspace}
\newcommand{\Xis}          {\ensuremath{\Xi^{\pm}}\xspace}
\newcommand{\Oms}          {\ensuremath{\Omega^{\pm}}\xspace}
\newcommand{\degree}       {\ensuremath{^{\rm o}}\xspace}

\begin{titlepage}
\PHyear{2022}       
\PHnumber{186}      
\PHdate{07 September}  

\title{Neutron emission in ultraperipheral Pb--Pb~collisions at $\mathbf{\sqrt{s_{\mathrm{NN}}}=5.02}$~TeV}
\ShortTitle{Neutron emission in UPC Pb–Pb at $\mathrm{\sqrt{s_{\mathrm{NN}}}=5.02}$~TeV}   

\Collaboration{ALICE Collaboration\thanks{See Appendix~\ref{app:collab} for the list of collaboration members}}
\ShortAuthor{ALICE Collaboration} 

\begin{abstract}
In ultraperipheral collisions (UPCs) of relativistic nuclei without overlap of nuclear densities, the two nuclei are excited by the Lorentz-contracted Coulomb fields of their collision partners. In these UPCs, the typical nuclear excitation energy is below a few tens of MeV, and a small number of nucleons are emitted in electromagnetic dissociation (EMD) of primary nuclei, in contrast to complete nuclear fragmentation in hadronic interactions. The cross sections of emission of given numbers of neutrons in UPCs of $^{208}$Pb nuclei at $\sqrt{s_{\mathrm{NN}}}=5.02$~TeV were measured with the neutron zero degree calorimeters (ZDCs) of the ALICE detector at the LHC, exploiting a similar technique to that used in previous studies performed at $\sqrt{s_{\mathrm{NN}}}=2.76$~TeV. In addition, the cross sections for the exclusive emission of one, two, three, four, and five forward neutrons in the EMD, not accompanied by the emission of forward protons, and thus mostly corresponding to the production of $^{207,206,205,204,203}$Pb, respectively, were measured for the first time. The predictions from the available models describe the measured cross sections well. These cross sections can be used for evaluating the impact of secondary nuclei on the LHC components, in particular, on superconducting magnets, and also provide useful input for the design of the Future Circular Collider (FCC-hh).
\end{abstract}
\end{titlepage}

\setcounter{page}{2} 


\section{Introduction} 

Studies of collisions of ultrarelativistic nuclei typically focus on the participant zone where the nuclei overlap and the quark--gluon plasma (QGP), a state of matter where quarks and gluons are free, can be created~\cite{Roland2014}. This is the domain where hot and dense matter is produced due to the enormous collision energy of the participating nucleons. At the same time, a domain of cold nuclear matter is expected to coexist with the participant zone in peripheral collisions. It is represented by spectator nucleons, which mainly preserve the velocity of the initial nuclei and travel forward. Measurements at the CERN Super Proton Synchrotron (SPS)~\cite{Appelshauser1998}  have shown that some of the spectator nucleons remain bound in nuclear fragments produced in peripheral collisions of ultrarelativistic $^{208}$Pb nuclei with a beam energy of 158~GeV per nucleon with a lead target. At this energy heavy spectator fragments were also detected in interactions of $^{208}$Pb nuclei with various target nuclei~\cite{Scheidenberger2004}. This was explained by a modest excitation of spectator matter in peripheral collisions~\cite{Scheidenberger2004}. 

In ultraperipheral collisions (UPCs) of relativistic heavy ions without nuclear overlap the colliding nuclei can be considered entirely as spectators. While their geometrical overlap is excluded because the collision impact parameter exceeds the sum of the nuclear radii, nuclei are still excited by the  Lorentz-contracted Coulomb fields of their collision partners. In comparison to hadronic nucleus--nucleus collisions, electromagnetic excitation is a rather soft process that results in the electromagnetic dissociation (EMD) of primary nuclei with the emission of just a few  nucleons~\cite{Pshenichnov2011}. 

Large EMD cross sections for the emission of one, two and three forward neutrons have been reported by the ALICE Collaboration~\cite{Abelev2012n} for UPCs of $^{208}$Pb nuclei at a centre-of-mass energy per nucleon pair $\sqrt{s_{\mathrm{NN}}}=2.76$~TeV at the Large Hadron Collider (LHC). EMD products retain beam rapidity similarly to spectators from hadronic nucleus--nucleus collisions. The EMD of $^{208}$Pb nuclei contributes significantly to the decay of the beam intensity~\cite{Baltz1996} and produces various secondary nuclei in all four interaction points (ALICE, ATLAS, CMS and LHCb) of the LHC~\cite{Bruce2009}. Certain secondary nuclei with their charge-to-mass ratio close to $^{208}$Pb (principally $^{207}$Pb) can travel for long distances around the collider rings because their magnetic rigidity is close to that of $^{208}$Pb and they may potentially be lost in superconducting magnets~\cite{Bruce2009,Hermes2016}. With a proper set-up of the collimation system, most of these secondary nuclei are intercepted efficiently by the off-momentum collimators and the dispersion suppressor sections around the LHC experiments. A dominant source of beam loss in the latter regions is the process of bound-free $\mathrm{e}^+\mathrm{e}^-$ pair production in UPCs with the electron capture by $^{208}$Pb~\cite{Bruce2009,Hermes2016}.  

The methods to study fragmentation of projectile nuclei at multi-GeV or TeV collision energies differ significantly from those used at much lower collision energies. In the latter case, fragments of projectiles with an initial energy of hundreds of MeV per nucleon are emitted at large angles with respect to the beam axis, so that their spread permits complete identification of projectile fragments using, among others, scintillator hodoscopes and time-of-flight  detectors~\cite{Botvina1995,Schuttauf1996}. In particular, the multi-fragment  break-up of spectator matter has been investigated at intermediate collision energies~\cite{Botvina1995}.

In contrast, the detection of projectile fragments in experiments at ultrarelativistic energies exceeding tens of GeV per nucleon is harder. It is difficult to separate these fragments from beam nuclei because they are emitted at very small angles with respect to the beam axis.  For example, projectile fragments were intercepted by imposing an external magnetic field and placing fragment detectors far from the interaction point~\cite{Appelshauser1998}. High resolution nuclear emulsions~\cite{Deines-Jones2000} and solid-state track detectors made of CR39 plastic~\cite{Cecchini2002} have also been employed to study fragmentation of ultrarelativistic nuclei. The charge distributions of fragments have been measured~\cite{Deines-Jones2000,Cecchini2002} in a wide range of charges of secondary nuclei, but nuclear emulsions and stacks of CR39 remained insensitive to spectator neutrons. Measurements of the charge-changing cross sections for lead and indium nuclei at an energy of 158~GeV per nucleon were performed with multiple-sampling ionisation chambers in Refs.~\cite{Scheidenberger2004} and~\cite{Uggerhjoj:In}, respectively, also without detecting forward neutrons. 

It should be emphasised that, to date, the production of charged spectator fragments has not been studied, neither at RHIC nor at the LHC. There exist only two proposals for future experiments to detect charged fragments in collider experiments, in particular, by means of a centrality detector at RHIC~\cite{Tarafdar2014} and by the ATLAS Forward Proton detectors at the LHC~\cite{Grinstein2016}. The ALICE experiment already has a unique possibility to detect forward protons as well as neutrons with its neutron and proton zero degree calorimeters (ZDCs)~\cite{Puddu2007,Gemme2009,Oppedisano2009}. 

According to the RELDIS ~\cite{Pshenichnov1999,Pshenichnov2001,Pshenichnov2011} and FLUKA~\cite{Braun2014} models, the production of various secondary nuclei is expected in EMD of $^{208}$Pb at the LHC.  It is impossible to identify masses and charges of  secondary nuclei at the LHC  and thus directly measure the cross sections of production of specific nuclides. However, the cross sections of production of $^{207,206,205,204,203}$Pb are closely related  to the cross sections to emit one, two, three, four, and five neutrons, respectively, in the absence of proton emission.

The aim of the present work is twofold. Firstly, to determine the cross sections of neutron emission in EMD of heavy nuclei at the highest collision energy available in accelerator experiments. Various models~\cite{Pshenichnov2011,Braun2014,Klusek-Gawenda2014,Broz2020} that are used to calculate neutron emission in EMD can be tested with these new data. Secondly, to measure the cross sections of emission of 1, 2, ..., 5 neutrons without protons,  to be used as approximations of the cross sections of production of  $^{207,206,205,204,203}$Pb. The validity of this approximation is studied by means of the RELDIS model.

In Sec.~\ref{Sec:Setup} the ALICE ZDCs are briefly described. In Sec.~\ref{Sec:ZED} the data sample and the adopted trigger configuration are presented. Section~\ref{Sec:Analysis} describes the methods of fitting ZDC energy spectra, and the corrections for detection efficiency and acceptance that provided the measured cross sections presented in Sec.~\ref{Sec:Results}.
In Sec.~\ref{Sec:discussion}, on the basis of the RELDIS model, the neutron emission without protons in EMD of $^{208}$Pb is associated with the production of secondary lead nuclei.  Finally, conclusions and outlook are given in Sec.~\ref{Sec:Conclusions}.

\section{Experimental set-up}\label{Sec:Setup}

\begin{figure}[!htb]
\begin{centering}
\includegraphics[width=1.0\columnwidth]{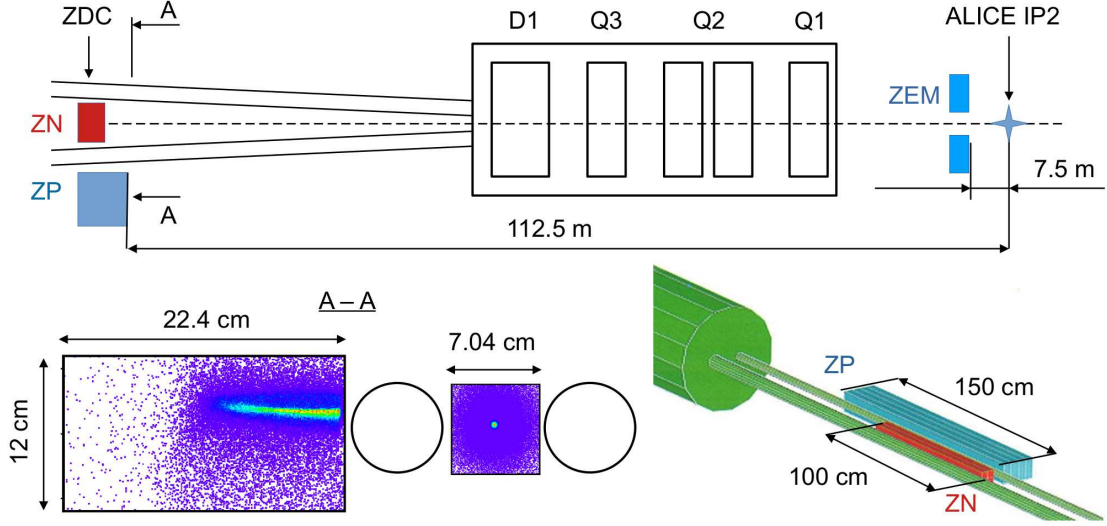}
\caption{Colour online: Simplified layout with respect to the ALICE interaction point (IP2) (not to scale) of the neutron (ZN), proton (ZP), and two electromagnetic (ZEM) calorimeters. The approximate positions of a dipole magnet (D1) and quadruple magnets (Q1, Q2, and Q3) are also shown. A view towards the forward surface of the ZN and ZP (A--A) shows typical distributions of entry points of forward nucleons obtained in Monte Carlo modelling. The longitudinal dimensions of ZN and ZP are shown in a simplified three-dimensional (3D) scheme.}
\label{fig:ZDC-ZEM-Layout}
\end{centering}
\end{figure}

A detailed description of the ALICE experiment can be found in
Ref.~\cite{ALICE2008JINST}. In the following, only the detectors relevant for the measurements discussed in this paper will be described. Two identical systems of hadronic calorimeters are placed on both sides (C and A) of the nominal interaction point (IP), 112.5~m away from the IP along the beam direction, see Fig.~\ref{fig:ZDC-ZEM-Layout} for the placement on the side A. The neutron (ZNC and ZNA) and proton (ZPC and ZPA) calorimeters~\cite{Puddu2007,Gemme2009,Oppedisano2009} were denoted as ZN and ZP, respectively.  The letter C(A) was assigned to the corresponding calorimeter because it intercepts the forward nucleons emitted by nuclei of the clockwise (anticlockwise) beam 1(2). Each ZDC is segmented into four towers. Half the optical fibres uniformly distributed in the calorimeter are read out by four tower photomultipliers (PMTs) and the other half are read out by a single fifth photomultiplier (PMC) common to all towers. Thus, there exist two options to obtain the energy deposited in the ZDC, either from the sum of signals in all PMTs (including PMC), or from the signal in the PMC alone. 

ZNC and ZNA are placed at zero degrees with respect to the neutron flight path from the IP to detect neutral forward particles at pseudorapidities $|\eta| > 8.8$. ZPC and ZPA  detect forward protons guided to these calorimeters by the LHC magnet system. Two small electromagnetic calorimeters (ZEM1 and ZEM2) are placed only on the side A, at 7.5 m from the IP, see Fig.~\ref{fig:ZDC-ZEM-Layout}, covering the pseudorapidity range $4.8\leq \eta \leq 5.7$~\cite{Oppedisano2009} and two intervals  $-16^\circ<\phi<16^\circ$ and $164^\circ<\phi <196^\circ$ of azimuthal angle. Technical characteristics of the hadronic neutron (ZN), proton (ZP), and electromagnetic (ZEM) calorimeters are summarised in Table~\ref{tab:ZDCTechSpec}.
\begin{table}[!tbh]
\caption{Technical characteristics of the hadronic neutron (ZN), proton (ZP), and electromagnetic (ZEM) calorimeters.}
\label{tab:ZDCTechSpec}
\begin{center}
\begin{tabular}{|c|c|c|c|}
\hline
 & ZN & ZP & ZEM \\
 \hline
Dimensions (cm$^3$)& $7.04\times 7.04\times 100$ & $12\times 22.4\times 150$ & $7\times 7\times 20.6$ \\
\hline
Absorber material &  W alloy & brass & lead  \\
\hline
$\rho_{\mathrm{absorber}}$ (g/cm$^3$) & 17.61 & 8.48 & 11.34 \\
\hline
Length (in $\lambda_\mathrm{I}$ units ) & 8.7 & 8.2 & 1.1 \\
\hline 
Length (in $X_\mathrm{0}$ units ) & 251 & 100 & 35.4 \\
\hline
Filling ratio & 1/22 & 1/65 & 1/22 \\
\hline 
Fibre spacing (mm) & 1.6 & 4 & -- \\
\hline
Fibre diameter (mm) & 0.365  & 0.550 & 0.550 \\
\hline
Fibre tilted at (deg) & 0 & 0 & 45 \\
\hline
\end{tabular}
\end{center}
\end{table}

\section{Data sample}\label{Sec:ZED}

Data on $^{208}$Pb--$^{208}$Pb collisions at $\sqrt{s_{\mathrm{NN}}}=5.02$~TeV collected by ALICE in 2018 were analysed. Special runs for the EMD cross section measurement with reduced instantaneous luminosity and, consequently, with reduced event pile-up were considered. For these runs the average number of hadronic inelastic interactions per bunch crossing $\mu_{\mathrm{inel}} \sim 1.3\times 10^{-4}$ was about 10 times lower than during the ALICE standard physics runs. Events were triggered requiring a signal over threshold either in ZNC or in ZNA. This condition is referred to as the ZED trigger and it is sensitive to one-sided EMD events with neutrons emitted either towards the C or A sides, as well as to EMD and hadronic events with neutrons emitted on both sides. It is customary (see, e.g.\ Ref.~\cite{Abelev2012n}) to define as single EMD events those with at least one neutron emitted in EMD processes, and as mutual EMD events those with at least one neutron emitted per side in EMD processes, so that mutual EMD events are a subset of single EMD events.

Using the ZDC timing information, only events corresponding to interactions happening at the nominal bunch crossing were selected, thus rejecting collisions between the beam and residual gas in the beam pipe, and collisions involving nuclei circulating outside the nominal bunch slots. The total number of selected events amounts to $ {N_{\mathrm{tot}}} = 2.050 \times 10^{6}$. The ZED trigger cross section $\sigma_{\mathrm{ZED}}$ was measured in a van der Meer (vdM) scan~\cite{vanderMeer1968} and found to be $\sigma_{\mathrm{ZED}} =  420.5 \pm 10.1$~b~\cite{Castellanos2021}.

The triggered events were classified into electromagnetic and hadronic events according to the signals of the forward  calorimeters ZEM1 and ZEM2. Events with a signal in either ZEM were tagged as hadronic and rejected, leaving a sample of ${N^{\mathrm{EMD}}}_{\mathrm{tot}} = 2.009 \times 10^{6}$ single or mutual EMD event candidates. However, as follows from the modelling of particle emission in UPCs of $^{208}$Pb at the LHC~\cite{Pshenichnov1999}, in a small fraction of EMD events, protons and pions are emitted at $|\eta| \leq 6$. Detection of these particles by one of the ZEMs may lead to the rejection of EMD events  due to the ZEM veto condition. Therefore, it is necessary to estimate the efficiency of the ZEM veto to select electromagnetic events. 

Let the numbers of  single EMD, mutual EMD and hadronic events be denoted as $s$, $m$ and $h$, respectively. The number of detected events with neutrons, say, only on side C is defined as $s-m+h\zeta_{hnm}={\cal N}_1$. In this equation a small fraction $\zeta_{hnm}$ of hadronic non-mutual events with the emission of forward neutrons only on one side is taken into account.  The number of events  with neutrons only on side C, but obtained without signal in ZEM is calculated as $s\varepsilon-m\varepsilon_m+h\zeta_{hnm}\varepsilon_h = {\cal N}_2$. Here the fractions of single EMD, mutual EMD, and hadronic events, which survive the ZEM veto, are denoted as $\varepsilon$, $\varepsilon_m$, and $\varepsilon_h$, respectively.  The condition that neutrons are detected only on side C can be released, thus providing the equations: $s+h = {\cal N}_3$ and $s\varepsilon + h\varepsilon_h ={\cal N}_4$, without and with the ZEM veto, respectively.
In Ref.~\cite{Abelev2012n} a small value $\varepsilon_h= 0.076$ was reported on the basis of HIJING modelling. According to the AAMCC model~\cite{Svetlichnyi2020}, the fraction of hadronic events with 1--5 forward neutrons on a given side, but without neutrons on the other side, is calculated as  $\zeta_{hnm}= 0.003$.  As estimated with the RELDIS and AAMCC models, for each of the 1n--5n channels, the relation $s \gg h$ is valid, and consequently, $s \gg h\varepsilon_h$. Because of the dominance of single EMD events over mutual EMD events, the relations $s-m \gg h\zeta_{hnm}$ and 
$s\varepsilon-m\varepsilon_m \gg h\zeta_{hnm}$ are also valid and the hadronic contribution can be neglected in all four equations for each neutron multiplicity with the solution $\varepsilon = {\cal N}_4/{\cal N}_3$ and $\varepsilon_m = ({\cal N}_4-{\cal N}_2)/({\cal N}_3-{\cal N}_1)$.

The resulting $\varepsilon_i$ for each neutron multiplicity $i=1,2,\dots 5$ for events with less than 6 neutrons and for events with any number of neutrons (Xn) are given in Table~\ref{tab:ZEM_eff_epsilon} together with their uncertainties calculated following  Ref.~\cite{Casadei:2009ic} assuming a uniform prior. The values of $\varepsilon_i$ estimated on both sides exceed 99\%. As can be also seen from Table~\ref{tab:ZEM_eff_epsilon}, there is a trend that the higher the neutron multiplicity, the higher the fraction of EMD events lost due to the ZEM veto. Hereafter, events with zero, one, two, ..., six neutrons are denoted as 0n, 1n, 2n, ..., 6n events.  Less than 0.125\% of 1n events are lost due to the ZEM veto, but up to 1\% for 5n. This is explained by the fact that EMD products associated with high neutron multiplicity events are emitted at larger angles and in some rare cases can hit ZEM. Nevertheless, the efficiency of selecting EMD events by imposing the ZEM veto remains extremely high ($>99.8$\%) for the 1n--5n channels. The values of $\varepsilon_i$ listed in Table~\ref{tab:ZEM_eff_epsilon} were used to correct the numbers of detected events ${n_i}$, of each neutron multiplicity $i$n, for the efficiency of the ZEM veto to select EMD events.
\begin{table}[!t]
\caption{Efficiency of the ZEM veto to select electromagnetic events with a given neutron multiplicity and for events with any number of neutrons, estimated from the data collected with and without the ZEM veto for the sides C and A.}
\label{tab:ZEM_eff_epsilon}
\begin{center}
\begin{tabular}{|c|c|c|}
\hline
Neutron & \multicolumn{2}{c|} {$\varepsilon_i$ (\%) } \\
 \cline{2-3}
multiplicity $i$n  & {Side C} & {Side A} \\
\hline
1n  & $99.875 \pm 0.005$ & $99.902  \pm 0.005$  \\
\hline
2n  & $99.766 \pm 0.014$ & $99.819 \pm 0.013$ \\
\hline
3n  & $99.457 \pm 0.039$ & $99.349 \pm 0.042 $\\
\hline 
4n  & $99.479 \pm 0.043$ & $99.321  \pm 0.049 $\\
\hline
5n  & $99.368 \pm 0.050$ & $99.025  \pm 0.064$ \\
\hline 
total 1n--5n & $99.802  \pm 0.005$  & $99.806  \pm 0.005 $ \\
\hline
total Xn    & $96.722 \pm 0.017$  & $96.117  \pm 0.019 $\\
\hline
\end{tabular}
\end{center}
\end{table}

The emission of neutrons in the EMD of one or both colliding nuclei can be accompanied by two-photon interactions $\gamma\gamma\rightarrow \mathrm{e}^+\mathrm{e}^-$ in the same UPC event. This process was considered in Ref.~\cite{Baltz2009} for mutual EMD. By modelling with STARlight~\cite{Klein2017} dilepton production and neutron emission on one side, the energy distributions of $\mathrm{e}^+$ and $\mathrm{e}^-$ within the acceptance of the ZEM were calculated. The probability to obtain leptons with energy above the ZEM veto threshold of about 10 GeV was found to be negligible.

\section{Analysis}\label{Sec:Analysis}

\subsection{Collection, calibration and fit of ZDC energy spectra}\label{Sec:ZDCSpectra}

In the present work the ZDC energy spectra collected with the PMC were used to obtain the neutron emission cross sections. These spectra  were calibrated to satisfy as precisely as possible two main criteria: (1) the pedestal peak is centred at zero energy; (2) the distance between consecutive peaks amounts to the beam energy $E_0=2510$~GeV per nucleon. 

The numbers of detected events $n_i$ for each neutron multiplicity class $i$n were extracted by fitting the calibrated distributions of energy $E$ deposited in ZN using the $\chi^2$ fit method. The fitting procedures were validated with Run~1 data in Ref.~\cite{Abelev2012n}. The fitting function $F(E)$ is the sum of six Gaussians corresponding to up to six detected neutrons
\begin{equation}
F(E)=\sum_{i=1}^{6} G_i(E)=\sum_{i=1}^{6} \frac{n_i}{\sqrt{2\pi} \sigma_i}\ \mathrm{e}^{-\frac{(E-\mu_i)^2}{2\sigma_i^2}} .
\label{Eq:FitFunc}
\end{equation}

The Gaussian $G_i(E)$ represents the $i$-th peak and it is characterised by the mean value $\mu_i$, the standard deviation $\sigma_i$ and the normalisation constant $n_i$, which denotes the number of detected events with $i$ neutrons. Because of the large widths of the peaks corresponding to neutron multiplicities larger than 5 in the neutron spectra, it was not possible to identify unambiguously the numbers of events with six or more neutrons. Though the 6n peak is not well defined, it is still used to improve fit results for 1,~2,~...,~5 neutrons. The values of  $\mu_1$, $\sigma_1$, and $n_1$ for the first nucleon peak were considered as free parameters. Despite the expected exact correspondence of $\mu_1$ to $E_0$, some variations of $\mu_1$ in the course of the fitting procedure were allowed within a $\pm 10$\% deviation from $E_0$. In addition, the value of $\sigma_1$ was constrained to be between $0.1\times E_0$ and $0.5\times E_0$ to improve the fit quality. These conditions make it possible to account for imperfect ZDC calibration. The parameters of the Gaussians describing the two-, three-, four-, five-, and six-nucleon peaks were also restricted: $\mu_i$ varied within $\pm 20$\% around $i \times \mu_1$ while $\sigma_i$ varied from $\sigma_1$ to $\sqrt{i} \sigma_1$. The numbers of detected events {$n_i$} in each peak were introduced as free parameters of the fit. As described below, after taking into account the acceptance$\times$efficiency of the ZDCs, corrected numbers of events $N_i$ were finally obtained from $n_i$. 

\begin{figure}[t]
\begin{centering}
\includegraphics[width=1.0\columnwidth]{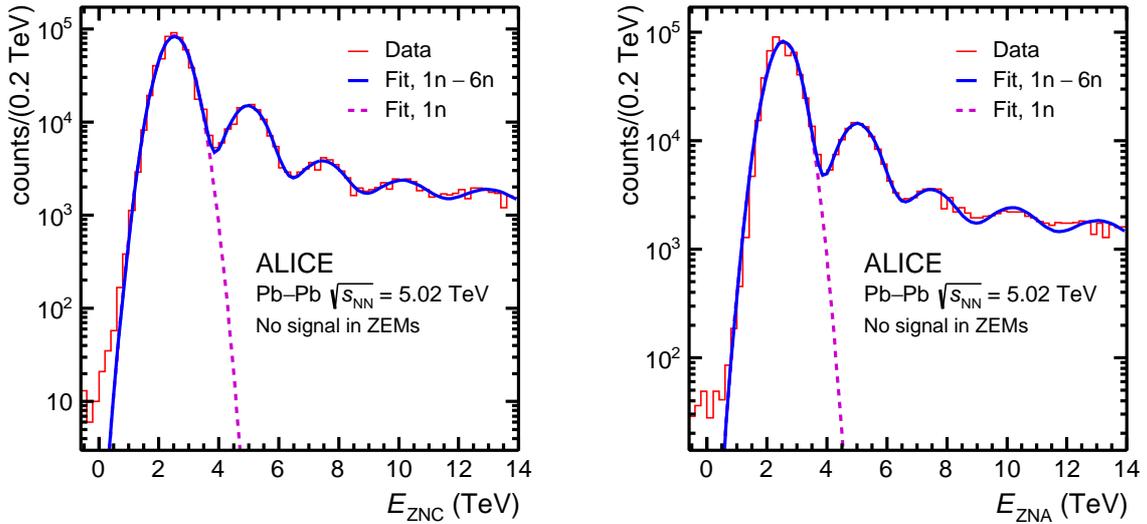}
\caption{Colour online: distributions of energy in ZNC (left) and ZNA (right) from EMD events (histograms) and resulting fit functions (solid curves) representing the sum of Gaussians. The Gaussians representing 1n peaks are shown by dashed curves.
}
\label{fig:Spectr_EM_n}
\end{centering}
\end{figure}

The distributions of energy deposited by forward EMD neutrons in ZNC and ZNA are shown in Fig.~\ref{fig:Spectr_EM_n}. These distributions were obtained for events with at least one neutron registered either in ZNC (Fig.~\ref{fig:Spectr_EM_n}, left) or in ZNA (Fig.~\ref{fig:Spectr_EM_n}, right), surviving the ZEM veto. As seen from this figure, the spectra collected on the C and A sides are very similar.  These spectra were fitted with the functions given by Eq.~(\ref{Eq:FitFunc}). The yields of  1n, 2n, ..., 5n EMD events were extracted from these spectra and used to calculate the cross section of emission of given numbers of neutrons along with any number of forward protons, as described in Sec.~\ref{Sec:n_p_mult}.

\subsection{Correction for detection efficiency and acceptance of neutron and proton ZDCs}
\label{Sec:corrections}
Some of the forward nucleons are lost on their way to the ZDCs due to the limited ZDC acceptance and scattering on various components of the LHC, in particular, on injection collimators, beam pipes, recombination chamber, and in air. Also, a peripheral impact of some nucleons on the calorimeter is responsible for an increased shower leakage leading to reduced energy deposited in the ZDCs. Because of such effects, the distributions of energy deposited in the ZDCs are distorted, particularly in events with a high nucleon multiplicity. For example, either one or two nucleons can be lost in a three-nucleon event. As a result, the three-nucleon event can be misidentified, respectively, as a two-nucleon or as a one-nucleon event. The probability to obtain a fake one-nucleon event will be different in two- and three-nucleon events.  This suggests the need to calculate the efficiency of event detection separately for each multiplicity class, as demonstrated in Ref.~\cite{Dmitrieva2018}.

The acceptance$\times$efficiency correction factors  $f_{i{\mathrm{n}}}=N^{\mathrm{MC}}_i/n^{\mathrm{MC}}_i$ were calculated for ZNC and ZNA and, for the study of neutron emission without protons, also for ZPC and ZPA. In a realistic Monte Carlo simulation of the ALICE apparatus, and after transporting neutrons generated with RELDIS, the number of generated events $N^{\mathrm{MC}}_i$ and the number of events $n^{\mathrm{MC}}_i$ registered in ZDC were calculated for each neutron multiplicity class $i$n. RELDIS has been validated with data on neutron emission in EMD of $^{208}$Pb at the LHC~\cite{Abelev2012n} and at lower collision energies~\cite{Golubeva2005}. The  acceptance$\times$efficiency correction factors calculated for the channels with the emission of zero, one, two, three, four, and five neutrons, possibly along with other particles, are given in Table~\ref{tab:n_reg_eff}. The values of $f_{i{\mathrm{n}}}$ were obtained by two different methods described below, and the simple averages $f_{i{\mathrm{n}}}$ were applied to raw data, as explained in Sec.~\ref{Sec:CrosUncert}. The first method was based on counting nucleons crossing the front area of each ZDC, while in the second method simulated ZDC energy spectra were fitted by the functions given by Eq.~(\ref{Eq:FitFunc}). Finally, in both methods the numbers of events of each multiplicity $i$n generated by RELDIS were divided by the numbers of detected events to obtain the correction factors. The difference between the results of the two methods divided by $\sqrt{2}$ is considered as the systematic uncertainty for the final average correction factors. These factors are given in Table~\ref{tab:n_reg_eff} together with their combined statistical and systematic uncertainties. As seen from this table, most of the obtained $f_{i{\mathrm{n}}}$ values exceed unity, because the loss of EMD neutrons in a channel of a certain multiplicity is  not compensated by the migration of events from less frequent channels of higher multiplicity induced by the loss of one or two neutrons.  In contrast, according to RELDIS, only in 3.2\% of EMD events neutrons are not emitted, but the 0n channel is filled with events of different neutron multiplicity when all neutrons are lost. This explains a noticeable correction of $f_{0{\mathrm{n}}}\approx 0.3$ for the 0n channel. 

When considering neutron emission without associated proton emission, additional correction factors for the proton ZDC, $f_{0{\mathrm p}}$, were applied to the raw yields. Hereafter,  events with zero protons are denoted as 0p events. These correction factors were calculated with the same two methods as for the neutron ZDC resulting in the average values of  $0.848 \pm 0.015$ and $0.852 \pm 0.018$  for ZPC and ZPA, respectively. In general, results from MC simulations are consistent with results of the probabilistic model of Ref.~\cite{Dmitrieva2018} with the probability  $p_{\mathrm n}=0.9$  to detect a neutron in ZN and with $p_{\mathrm p}=0.55$ to detect a proton in ZP. 
\begin{table}[t]
\caption{Acceptance$\times$efficiency correction factors for detecting neutrons in ZNC and ZNA and their estimated uncertainties.}
\label{tab:n_reg_eff}
\begin{center}
\begin{tabular}{|c|c|c|}
\hline
Neutron       & \multicolumn{2}{|c|}{Average of two methods} \\
multiplicity  & \multicolumn{2}{|c|}{$f_{i{\mathrm{n}}}$} \\
\cline{2-3} 
$i$n           &  ZNC & ZNA \\
\hline
0n &  0.286 $\pm$ 0.126 & 0.302 $\pm$ 0.097 \\
\hline
1n & 1.064 $\pm$ 0.031  & 1.064 $\pm$ 0.030 \\
\hline
2n & 1.092 $\pm$ 0.024  & 1.010 $\pm$ 0.095 \\
\hline
3n & 1.057 $\pm$ 0.032  & 1.066 $\pm$ 0.020 \\
\hline
4n & 1.001 $\pm$ 0.046  & 0.962 $\pm$ 0.094 \\
\hline
5n & 0.907 $\pm$ 0.132  & 0.917 $\pm$ 0.104 \\
\hline
\end{tabular}
\end{center}
\end{table}

\subsection{Determination of neutron emission cross sections}
\label{Sec:CrosUncert}

The cross sections for specific EMD channels $\sigma(i{\mathrm{n}})$ with a given number of neutrons $i$ and an arbitrary number of protons were obtained by combining $\sigma_{\mathrm{ZED}}$ introduced in Sec.~\ref{Sec:ZED} and ${n_i}$ from the fit functions given by Eq.~(\ref{Eq:FitFunc}). The cross sections were measured separately on the C and A sides by ZNC and ZNA, respectively. After the signal extraction, the corrections for the efficiency of the ZEM veto $\varepsilon_i$ introduced in Sec.~\ref{Sec:ZDCSpectra} and for the ZDC efficiency and acceptance, $f_{i{\mathrm{n}}}$, Sec.~\ref{Sec:corrections}, were applied:
\begin{equation}
\sigma(i{\mathrm{n}})= \sigma_{\mathrm{ZED}}\frac{N_i}{N_{\mathrm{tot}}} = \sigma_{\mathrm{ZED}} \frac{n_i}{N_{\mathrm{tot}}} \frac{f_{i{\mathrm{n}}}}{\varepsilon_i}=\sigma_{\mathrm{ZED}} \frac{n_i}{N_{\mathrm{tot}}}F_{i\mathrm{n}} \ .
\label{Eq:SigChanneln}
\end{equation}
Here $N_i$ is a corrected number of events with given neutron multiplicity $i$, $N_{\mathrm{tot}}$ is the total number of ZED trigger events defined in Sec.~\ref{Sec:ZED}, and $F_{i\mathrm{n}}$ is the resulting correction factor. The cross sections of specific EMD channels $\sigma(i{\mathrm{n}},{\mathrm{0p}})$ with given number of neutrons $i$ and without protons on the side C or A were obtained in the same way, but with the additional  correction factors $f_{0{\mathrm p}}$ for the efficiency of registration of 0p events in ZPC and ZPA, respectively:
\begin{equation}
\sigma(i{\mathrm{n}},\mathrm{0p})= \sigma_{\mathrm{ZED}}\frac{N_i}{N_{\mathrm{tot}}} = \sigma_{\mathrm{ZED}} \frac{n_i}{N_{\mathrm{tot}}} \frac{f_{i{\mathrm{n}}} f_{0{\mathrm p}}}{\varepsilon_i}= \sigma_{\mathrm{ZED}} \frac{n_i}{N_{\mathrm{tot}}}F_{i\mathrm{n},0\mathrm{p}} \ .
\label{Eq:SigChannelp}
\end{equation}

Here $F_{i\mathrm{n},0\mathrm{p}}$ is the resulting correction factor as in Eq.~(\ref{Eq:SigChanneln}), but for 0p events. Because of the very large number of collected ZED trigger events, the uncertainty of $N_{\mathrm{tot}}$ is negligible. The statistical uncertainties of $n_i$ originate from the uncertainties on the numbers of events found by the fit procedure. The systematic uncertainties of $n_i$ were estimated from a  variation of the fit procedure by considering the difference in $n_i$ obtained with the nominal and double bin size.  Only systematic uncertainties were considered for $F_{i\mathrm{n}}$ and $F_{i\mathrm{n},0\mathrm{p}}$. Following Ref.~\cite{Castellanos2021}, the relative systematic uncertainty of $\sigma_{\mathrm{ZED}}$ was taken as $2.4$\%, resulting from the
vdM scan analysis.

The final cross sections were obtained as the average between the measurements performed on the sides C and A from the event numbers and resulting correction factors introduced in Eqs.~(\ref{Eq:SigChanneln}) and (\ref{Eq:SigChannelp}):
\begin{equation}
\bar{\sigma}=\frac{\sigma^\mathrm{C}+\sigma^\mathrm{A}}{2}=\sigma_{\mathrm{ZED}}\frac{n^\mathrm{C} F^\mathrm{C} + n^\mathrm{A} F^\mathrm{A}}{2N_{\mathrm{tot}}} \ \ .
\label{Eq:Average}
\end{equation}

The contributions to the relative systematic uncertainties of $\bar{\sigma}$ were propagated from the systematic uncertainties on each side as summarised in Table~\ref{tab:syst_uncert}. These uncertainties were calculated separately for the cross sections of neutron emission accompanied by an arbitrary number of protons (Yp, including zero protons) and without proton emission (0p).
\begin{table}[t]
\caption{Relative systematic uncertainties of the cross sections of emission of given numbers of neutrons $i$ accompanied by an arbitrary number of protons (Yp, including zero protons) and without protons (0p), in UPCs of $^{208}$Pb nuclei at $\sqrt{s_{\mathrm{NN}}}=5.02$~TeV. Each uncertainty is calculated for the average of the cross sections measured on the sides C and A.}
\label{tab:syst_uncert}
\begin{center}
\begin{tabular}[c]{|c|c|c|c|c|c|}
\hline
\multirow{3}*{Source}	& \multicolumn{5}{c|}{Relative uncertainty  (\%)} \\
\cline{2-6}
	& {1n} & {2n} &{3n} &{4n} & {5n} \\
\cline{2-6}
	& Yp $\vert$ \ \rm{0p} \ & Yp $\vert$ \ \rm{0p} \ & Yp $\vert$ \ \rm{0p} \ & Yp $\vert$ \ \rm{0p} \ & Yp $\vert$ \ \rm{0p} \ \\
\hline
Fitting procedure   & 0.55 $\vert$ \ 0.55   & 0.32 $\vert$ \ 0.29   & 0.83 $\vert$ \ 0.72 & 0.73 $\vert$ \ 0.67     & 1.14 $\vert$ \ 1.01 \\
\hline
ZDC+ZEM efficiency      & 2.03 $\vert$ \ 2.45   & 4.68 $\vert$ \ 4.88   & 1.78 $\vert$ \ 2.25 & 5.35 $\vert$ \ 5.52   & 9.26 $\vert$ \ 9.36 \\
\hline
$\sigma_{\mathrm{ZED}}$ determination& \multicolumn{5}{c|}{\multirow{2}*{2.4}} \\ 
 from vdM scan & \multicolumn{5}{c|}{}\\
\hline
Total                   & 3.19 $\vert$ \ 3.47   & 5.27 $\vert$ \ 5.45 & 3.10 $\vert$ \ 3.37 & 5.91 $\vert$ \ 6.05    & 9.63 $\vert$ \ 9.72 \\
\hline
\end{tabular}
\end{center}
\end{table}

In these measurements, the pile-up of EMD events was lower with respect to the previous ALICE measurements of the EMD cross sections~\cite{Abelev2012n}. Nevertheless, the uncertainty due to the residual pile-up of EMD events was also addressed.  The main issue  with the pile-up is to count two 1n events as a 2n event. For these runs the average number of hadronic inelastic interactions per bunch crossing $\mu_{\mathrm{inel}}$ was about $1.3 \times 10^{-4}$. The same value, but calculated for 1n emission in EMD is $\mu_{\mathrm{1n}}=\mu_{\mathrm{inel}}\times (\sigma_{\mathrm{1n}} / \sigma_{\mathrm{had}})=1.8\times 10^{-3}$, where  $\sigma_{\mathrm{1n}} = 108$~b is calculated with RELDIS and $\sigma_{\mathrm{had}} = 7.67\pm 0.25$~b is the inelastic hadronic interaction cross section measured in Ref.~\cite{Castellanos2021}. With $\sigma_{\mathrm{2n}}/\sigma_{\mathrm{1n}}\approx 0.24$, also estimated with RELDIS, the relation $\mu_{\mathrm{2n}}\approx 0.24\times \mu_{\mathrm{1n}}$ is obtained. Following the Poisson distribution, the ratio of the probabilities to obtain two 1n events in the same bunch crossing $p(2,\mu_{\mathrm{1n}})$ and to have one 2n event $p(1,\mu_{\mathrm{2n}})$ is
\begin{equation}
\frac{p(2,\mu_{\mathrm{1n}})}{p(1,\mu_{\mathrm{2n}})}=\frac{\mu^2_{\mathrm{1n}}  \exp(-\mu_{\mathrm{1n}})}{2\mu_{\mathrm{2n}}  \exp(-\mu_{\mathrm{2n}})}=\frac{\mu_{\mathrm{1n}}}{0.48}\exp(-0.76\mu_{\mathrm{1n}})\approx 0.004 \ .
\end{equation}
One can conclude that the pile-up effect of 0.4\% for 2n emission can be neglected in view of the total $\approx 5$\% uncertainty for this channel reported in Table~\ref{tab:syst_uncert}.

\section{Results}\label{Sec:Results}

\subsection{Cross sections of emission of given numbers of neutrons}
\label{Sec:n_p_mult}

The EMD cross sections $\sigma(i{\mathrm{n}})$ for 1n, 2n, 3n, 4n, and 5n emission accompanied by an arbitrary number of forward protons (including zero protons) were measured separately on the C and A sides. They are listed in Table~\ref{tab:n_multiplicity}. In order to obtain these cross sections, EMD events were selected by applying a veto on ZEM, as described in Section~\ref{Sec:ZED}. The cross sections given in Table~\ref{tab:n_multiplicity} were corrected for the ZDC and ZEM efficiency separately for each side as described in Sec.~\ref{Sec:corrections}. As seen, the cross sections measured on the side C are slightly larger compared to those on the side A.  However, the side C and A cross sections are in most cases  consistent, because this difference is within the uncorrelated uncertainty of the side A cross section. As follows from Table~\ref{tab:syst_uncert}, the main contribution to the systematic uncertainties of the cross sections is due to the uncertainties of the corrections for ZDC and ZEM efficiency. 

The average $\sigma(i{\mathrm n})$ between the C and A sides were calculated for each neutron multiplicity $i$ according to Eq.~(\ref{Eq:Average}). The resulting $\sigma(i{\mathrm{n}})$ are given in Table~\ref{tab:n_multiplicity} with their statistical and systematic uncertainties propagated from the uncertainties on each side, which were considered as uncorrelated, except for the contribution from the vdM scan uncertainty, which is fully correlated between the C and A sides. The differences between the cross sections obtained on the sides C and A depend on the chosen method to calculate the ZDC efficiency, with the exception of 1n channel, so the uncertainties on the efficiency can justify the difference between the cross sections on the C and A sides. On the basis of this finding, the difference between C and A sides divided by $\sqrt{2}$ was considered as an additional uncertainty only for the average 1n cross section.

In Table~\ref{tab:n_multiplicity} the measured cross sections are compared with results of the  RELDIS~\cite{Pshenichnov2011} and $\mathrm{n^O_On}$~\cite{Broz2020} models. Both models are based on the Weizs\"{a}cker--Williams method to calculate the cross sections of neutron emission in EMD of nuclei by considering the respective photonuclear reactions induced by equivalent photons. RELDIS simulates nuclear reactions induced by Weizs\"{a}cker--Williams photons on $^{208}$Pb by means of the intranuclear cascade model of photonuclear reactions with the subsequent de-excitation of residual nuclei via neutron evaporation and other processes~\cite{Pshenichnov2005}. A phenomenological approximation for the total photoabsorption cross section on $^{208}$Pb is used as an input to RELDIS  together with calculated relative contributions of various final states to obtain the absolute values of $\sigma(i{\mathrm{n}})$ and $\sigma(i{\mathrm{n}},{\mathrm{0p}})$. A similar phenomenological approximation of the total photoabsorption cross section is also used in $\mathrm{n^O_On}$. In this event generator the calculations of probability of a given neutron emission channel are based on approximations of partial photoneutron cross sections measured on $^{208}$Pb below 140~MeV and extrapolated to higher photon energies~\cite{Broz2020}. Only neutron emission events can be generated with $\mathrm{n^O_On}$, but the 0n EMD cross section of 6.85~b  corresponding to 3.2\% of EMD events with the emission of other particles without neutrons was  calculated with RELDIS.     

The uncertainties of RELDIS results ($\approx 5$\%) given in Table~\ref{tab:n_multiplicity} stem from the uncertainty of the phenomenological approximation of the total photoabsorption cross section on $^{208}$Pb used by this model. As shown in Ref.~\cite{Pshenichnov2001}, the uncertainties of the calculations of specific neutron emission channels (1n, 2n) in absorption of low energy photons are typically as high as 7\%. Another source of uncertainties is connected to the estimations of unknown total nuclear photoabsorption cross section at high photon energies ($>60$~GeV) from the measured total $\gamma\mathrm{p}$ cross sections taking into account nuclear shadowing. As evaluated  recently~\cite{Pshenichnov2019}, the total EMD cross sections calculated with different high-energy approximations for $^{208}$Pb--$^{208}$Pb collisions at the LHC differ by $\approx 2$\%. Similar uncertainties of $\approx$~2--4\% were attributed to the measured total cross section of the absorption of real intermediate energy photons (0.5--2.6~GeV) by $^{\mathrm{nat}}$Pb~\cite{Muccifora1999}. The total EMD cross section is the sum of the contributions of processes induced by Weizs\"{a}cker--Williams  photons from the above-mentioned energy domains. Therefore, the combined error of $\approx 5$\% is attributed to the cross sections calculated with RELDIS also for each individual neutron emission channel in Table~\ref{tab:n_multiplicity}. The uncertainty of 2\% also applies to the cross sections calculated with $\mathrm{n^O_On}$ because of the uncertainties of the approximation of the total photoabsorption cross section on $^{208}$Pb. 

The sum of the measured 1n--5n cross sections amounts to $151.5\pm 0.2 \pm 4.7$~b, and it is within $2\sigma$ in agreement with the same sum calculated with RELDIS as $159.8\pm 5.6$~b, see Table~\ref{tab:n_multiplicity}. The 1n--5n sum calculated with $\mathrm{n^O_On}$ as $143.1 \pm 2.2$~b is lower. In general, the measured sum is in the middle between the models. The total single EMD cross section for neutron emission can be simply estimated from the cross sections measured in Ref.~\cite{Castellanos2021} as  $(\sigma_{\mathrm{ZED}}-\sigma_{\mathrm{had}})/2=206.4$~b assuming 100\% efficiency of the ZED trigger. It is in good agreement with the corresponding cross section of $204.6\pm 7.1$~b  calculated with RELDIS. As seen, the measurements and calculations demonstrate dominant contributions of 73\% and 78\%, respectively, of 1n--5n channels to the EMD of $^{208}$Pb leading to neutron emission.

\begin{table}[tb]
\caption{Cross sections of emission of one, two, three, four and five neutrons along with an arbitrary number of protons in EMD of $^{208}$Pb at $\sqrt{s_{\mathrm{NN}}} = 5.02$~TeV measured on the C and A sides and their average along with their statistical and systematic uncertainties. Cross sections calculated with RELDIS~\cite{Pshenichnov2011} and $\mathrm{n^O_On}$~\cite{Broz2020}  are given for comparison. The sum of 1n--5n cross sections is presented in the last row. The uncertainties of the measurements are reported in the order $\pm$ (stat.)  $\pm$ (syst.). }
\label{tab:n_multiplicity}
\begin{center}
\begin{tabular}{|c|l|l|l|l|l|}
\hline
\multirow{2}*{ZN} & \multicolumn{2}{c|}{$\sigma(i{\mathrm{n}})$ (b)}& \multicolumn{1}{c|}{$\sigma(i{\mathrm{n}})$}& \multicolumn{1}{c|}{$\sigma^{\mathrm{RELDIS}}(i{\mathrm{n}})$} & \multicolumn{1}{c|}{$\sigma^{\mathrm{n^O_On}}(i{\mathrm{n}})$} \\
\cline{2-3}
& \multicolumn{1}{c|}{Side C} & \multicolumn{1}{c|}{Side A} &\multicolumn{1}{c|}{(b)} & \multicolumn{1}{c|}{(b)} & \multicolumn{1}{c|}{(b)}\\
\hline
1n &  $ 109.7 \pm 0.1 \pm 4.2 $  & $ 107.2 \pm 0.1 \pm 4.1 $  & $ 108.4 \pm 0.1  \pm 3.9 $  & $108.0 \pm 5.4$ & $103.7 \pm 2.1$\\
\hline
2n &  $ 25.8 \pm 0.1 \pm 0.8 $ & $ 24.1 \pm 0.1 \pm 2.3 $   & $ 25.0 \pm 0.1  \pm 1.3$   & $25.9 \pm 1.3$  & $23.6 \pm 0.5$ \\
\hline
3n &  $ 7.97 \pm 0.07  \pm 0.33 $  & $ 7.94 \pm 0.04 \pm 0.25 $    & $ 7.95 \pm 0.04  \pm 0.25$  & $11.4 \pm 0.6$  & $6.3 \pm 0.1$ \\
\hline
4n &  $ 5.73 \pm 0.04 \pm 0.30 $    & $ 5.56 \pm 0.04 \pm 0.56 $    & $ 5.65 \pm 0.03 \pm 0.33$  & $7.8 \pm 0.4$   & $4.8 \pm 0.1$ \\
\hline
5n &  $ 4.61 \pm 0.04  \pm 0.68 $      & $ 4.47 \pm 0.04 \pm 0.52 $ & $ 4.54 \pm 0.03 \pm 0.44$  & $6.3 \pm 0.3$   & $4.7 \pm 0.1$ \\
\hline
1n--5n &  &  & $ 151.5 \pm 0.2 \pm 4.7$  & $159.8 \pm 5.6$   & $143.1 \pm 2.2$ \\
\hline
\end{tabular}
\end{center}
\end{table}

\begin{figure}[t]
\begin{centering}
\includegraphics[width=0.85\columnwidth]{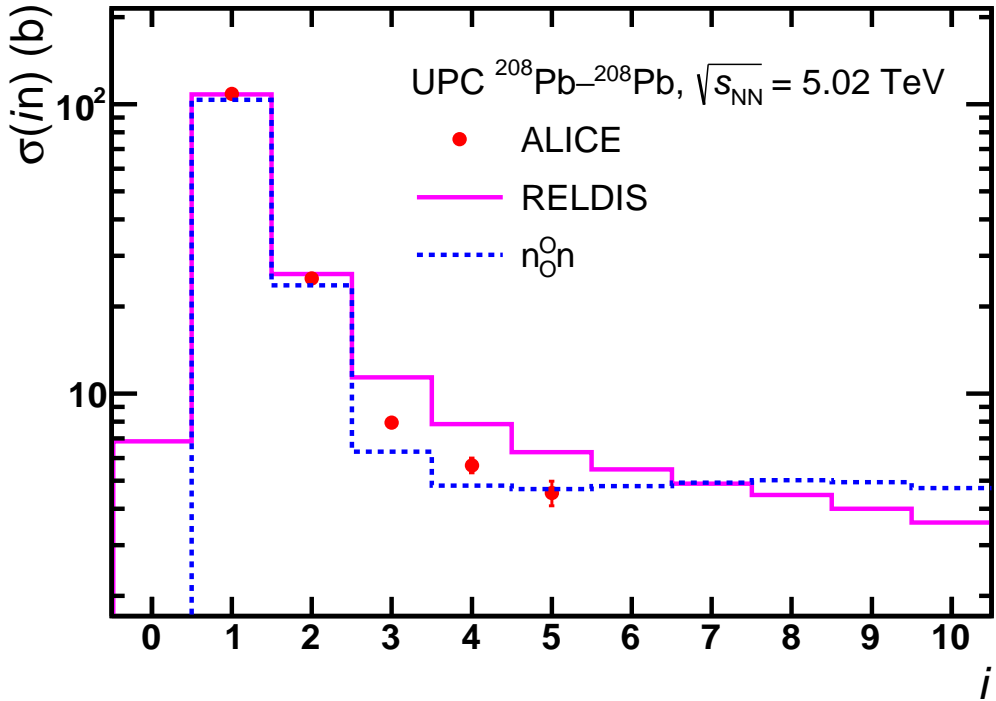}
\caption{Colour online: measured (points) and calculated with RELDIS~\cite{Pshenichnov2011} (solid-line histogram) and $\mathrm{n^O_On}$~\cite{Broz2020} (dashed-line histogram) cross sections of emission of given numbers of neutrons $i$ in UPCs of $^{208}$Pb nuclei at $\sqrt{s_{\mathrm{NN}}}=5.02$~TeV. Combined statistical and systematic uncertainties of the measurements are presented.}  
\label{fig:neutrons}
\end{centering}
\end{figure}

The measured cross sections are also shown in Fig.~\ref{fig:neutrons} together with results of the models.  As seen, the cross sections of 1n- and 2n-emission calculated  with RELDIS and $\mathrm{n^O_On}$ agree with the measured ones. On the other hand, the measured 3n and 4n cross sections are overestimated by RELDIS and underestimated by $\mathrm{n^O_On}$. At the same time, the 5n cross section is described very well by $\mathrm{n^O_On}$, but overestimated by RELDIS. One can note that the models agree well with respect to the 1n and 2n cross sections, but differ for higher neutron multiplicities. This indicates the importance of 3n, 4n, and 5n measurements for tuning the parameters of the models mentioned above.

\subsection{Comparison with previous ALICE results}
\label{Sec:comparison}

Previous ALICE results on neutron emission in EMD were reported in Ref.~\cite{Abelev2012n} for $^{208}$Pb--$^{208}$Pb collisions at $\sqrt{s_{\mathrm{NN}}}=2.76$~TeV. Events were required to have a signal in one of the two ZN and no signal in the other one, so events of symmetric emission, such as those occurring in hadronic and mutual EMD were rejected, and the sample contained single but not mutual EMD events.  The ZEM veto was not used to suppress hadronic events. With this selection the fractions of 1n, 2n, and 3n events were determined in Ref.~\cite{Abelev2012n}. Following the same selection, the spectra in ZNC and ZNA for single minus mutual EMD process were obtained also in the present work at $\sqrt{s_{\mathrm{NN}}}=5.02$~TeV without the ZEM veto described in Sec.~\ref{Sec:ZED}.

The fractions of 1n-, 2n-, 3n-, and 4n-events over the total number of events without neutrons on the opposite side $N_{\mathrm{one\mbox{-}side}}$ are listed in Table~\ref{tab:Neutron_fractions}. These fractions  were extracted from ZNC and ZNA spectra and were corrected for ZNC and ZNA acceptance as described in Sec.~\ref{Sec:corrections}. The measurements are compared with RELDIS results given in the same table. The previous ALICE results for $^{208}$Pb--$^{208}$Pb~collisions at $\sqrt{s_{\mathrm{NN}}}=2.76$~TeV~\cite{Abelev2012n} are also given together with RELDIS results for this collision energy.

\begin{table}[t]
\caption{Fractions (in \%) of EMD events with given numbers of neutrons but without neutrons on the  opposite side measured and calculated for $^{208}$Pb--$^{208}$Pb UPCs at $\sqrt{s_{\mathrm{NN}}}=2.76$~TeV~\cite{Abelev2012n} and $\sqrt{s_{\mathrm{NN}}}=5.02$~TeV. The uncertainties of the measurements are given in the order $\pm$ (stat.)  $\pm$ (syst.).}
\label{tab:Neutron_fractions}
\begin{center}
\begin{tabular}{|c|c|c|c|c|}
\hline
 Single minus & \multicolumn{2}{c|}{$\sqrt{s_{\mathrm{NN}}}=2.76$~TeV~\cite{Abelev2012n}} & \multicolumn{2}{c|}{$\sqrt{s_{\mathrm{NN}}}=5.02$~TeV} \\
\cline{2-5}
mutual &  Experiment &  RELDIS  &  Experiment  & RELDIS  \\
\hline
${N_1}/{N_{\mathrm{one\mbox{-}side}}}$ & $51.5 \pm 0.4 \pm 0.2$ & $54.2 \pm 2.4$ & $52.4 \pm 0.1 \pm 1.3 $ \ \ \ \ \ & $51.3 \pm 2.3$ \\
\hline
${N_2}/{N_{\mathrm{one\mbox{-}side}}}$ & $11.6 \pm 0.3 \pm 0.5$ & $12.7 \pm 0.8$ & $11.94 \pm 0.03  \pm 0.56 $ & $12.2 \pm 0.8$ \\
\hline
${N_3}/{N_{\mathrm{one\mbox{-}side}}}$ & $3.6 \pm 0.2 \pm 0.2$ & $5.4 \pm 0.7$ & $3.74 \pm 0.02 \pm 0.07 $  & $5.4 \pm 0.7$ \\
\hline 
${N_4}/{N_{\mathrm{one\mbox{-}side}}}$ & & & $2.66 \pm 0.01 \pm 0.14 $  & $3.7 \pm 0.5$ \\
\hline
\end{tabular}
\end{center}
\end{table}

As seen from Table~\ref{tab:Neutron_fractions}, similar fractions of 1n, 2n, and 3n events are measured at $\sqrt{s_{\mathrm{NN}}}=2.76$~TeV~\cite{Abelev2012n} and $\sqrt{s_{\mathrm{NN}}}=5.02$~TeV. The dominance of 1n emission is evident at both collision energies. The measured 1n and 2n fractions are in very good agreement with RELDIS. A slight reduction of calculated 1n events is seen at higher energy because of the redistribution of events in favour of higher multiplicities. This can be explained by a higher average equivalent photon energy at $\sqrt{s_{\mathrm{NN}}}=5.02$~TeV. However, such a subtle effect cannot be traced with confidence because of measurement and modelling uncertainties, which were calculated as described in Sec.~\ref{Sec:CrosUncert}.


\subsection{Emission of neutrons without protons}
\label{Sec:n_0p_mult}

In order to measure the cross section for 1n, 2n, ..., 5n emission without proton emission (0p), the energy spectra in ZNC and ZNA were obtained for events not having a signal in the respective ZP. Measured cross sections of 1n-, 2n-, 3n-, 4n-, and 5n-emission without protons are given in Table~\ref{tab:Pb_production} together with RELDIS results.  As can be seen, the cross sections measured on side C are systematically slightly larger than on side A, but the difference remains within the uncorrelated uncertainty of side A cross sections for all neutron multiplicities.  The cross sections were calculated with Eq.~(\ref{Eq:SigChannelp}) from the numbers of events ${n_i}$ of each neutron multiplicity $i$n obtained from the fit of spectra  with the functions given by Eq.~(\ref{Eq:FitFunc}). In this case of 0p measurements, the numbers of true events of each multiplicity ${N_i}$ were obtained  by correcting for the efficiency of ZNC and ZNA and also for protons undetected in ZPC and ZPA, as described in Sec.~\ref{Sec:corrections}. Statistical and systematic uncertainties were calculated as described in Sections~\ref{Sec:CrosUncert} and~\ref{Sec:n_p_mult}. Contributions to the total systematic uncertainties are presented in Table~\ref{tab:syst_uncert}. 

The measured cross sections of 1n, 2n, ..., 5n emission not accompanied by protons, shown in Table~\ref{tab:Pb_production}, are lower than the cross sections of 1n, 2n,.. 5n emission along with other particles presented in Table~\ref{tab:n_multiplicity}. This difference is larger for 4n and 5n channels because protons are emitted more frequently at higher equivalent photon energies associated with multineutron events. With the exception of 5n emission, the measured 0p cross  sections are lower than the cross sections calculated with  RELDIS. Nevertheless, the agreement between the measurements and calculations for 0p cross sections is better than for Yp cross sections presented in  Table~\ref{tab:n_multiplicity}.

The sum of the measured 0p cross sections from 1n to 5n is estimated as $126.0\pm 0.2\pm 4.1$~b. It is also listed in Table~\ref{tab:Pb_production} for comparison with RELDIS.
This sum of 0p cross sections provides a dominant ($\approx 83$\%) contribution to the sum of Yp cross sections $151.5\pm 0.2\pm 4.7$~b given in Table~\ref{tab:n_multiplicity}. In other words, only $\approx 17$\% of 1n--5n events are accompanied by the emission of protons. 

\begin{table}[t]
\caption{Cross sections of emission of one, two, three, four and five neutrons without protons on the same side in the EMD of $^{208}$Pb at $\sqrt{s_{\mathrm{NN}}} = 5.02$~TeV, measured on the C and A sides, and their average along with their statistical and systematic uncertainties. Cross sections calculated with RELDIS are given for comparison. The sum of 1n--5n cross sections is presented in the last row. The uncertainties of the measurements are given in the order $\pm$ (stat.) $\pm$ (syst.).}
\label{tab:Pb_production}
\begin{center}
\begin{tabular}{|c|c|c|c|c|c|}
\hline
\multirow{2}*{ZN} & \multirow{2}*{ZP} & \multicolumn{2}{c|}{$\sigma(i{\mathrm{n}},{\mathrm{0p}})$ (b)}& \multirow{2}*{ $\sigma(i{\mathrm{n}},{\mathrm{0p}})$ (b)}& \multirow{2}*{ $\sigma^{\mathrm{RELDIS}}(i{\mathrm{n}},{\mathrm{0p}})$ (b)}\\
\cline{3-4}
& &  Side C &  Side A &  & \\
\hline
1n & \multirow{5}*{0p} & $ 92.6 \pm 0.1  \pm 3.9 $ \ \ \ \ & $ 90.9 \pm 0.1  \pm 4.0 $ \ \ \ \ & $ 91.8 \pm 0.1 \pm 3.4 \ \ \ \ $   & $104.1 \pm 5.2$ \ \ \  \\
\cline{1-1}
\cline{3-6}
2n & & $ 21.4 \pm 0.1  \pm  0.8 $ \ \ \ \ & $ 20.0 \pm 0.1  \pm 2.0  $ \ \ \ \  & $ 20.7 \pm 0.1 \pm 1.1 $  \ \ \ \  & $21.9 \pm 1.1$ \ \ \ \  \\
\cline{1-1}
\cline{3-6}
3n & & $ 6.14 \pm 0.07 \pm 0.28 $  & $ 6.21 \pm 0.04 \pm 0.24 $  & $ 6.17 \pm 0.04 \pm 0.21 $  & $7.59 \pm 0.38$\\
\cline{1-1}
\cline{3-6}
4n & & $ 4.21 \pm 0.04 \pm 0.23 $  & $ 4.08 \pm 0.04 \pm 0.42 $   & $ 4.15 \pm 0.03 \pm 0.25 $  & $4.29 \pm 0.22$ \\
\cline{1-1}
\cline{3-6}
5n & & $ 3.16 \pm 0.04 \pm 0.47 $  & $ 3.08 \pm 0.03 \pm 0.36 $   & $ 3.12 \pm 0.03 \pm 0.30 $  & $2.95 \pm 0.15$ \\
\cline{1-1}
\cline{3-6}
1n--5n  &  &  &  & $ 126.0 \pm 0.2 \pm 4.1 $ \ \ \ & $140.8 \pm 5.3$ \ \ \ \\
\hline
\end{tabular}
\end{center}
\end{table}

\begin{figure}[t]
\begin{centering}
\includegraphics[width=0.85\columnwidth]{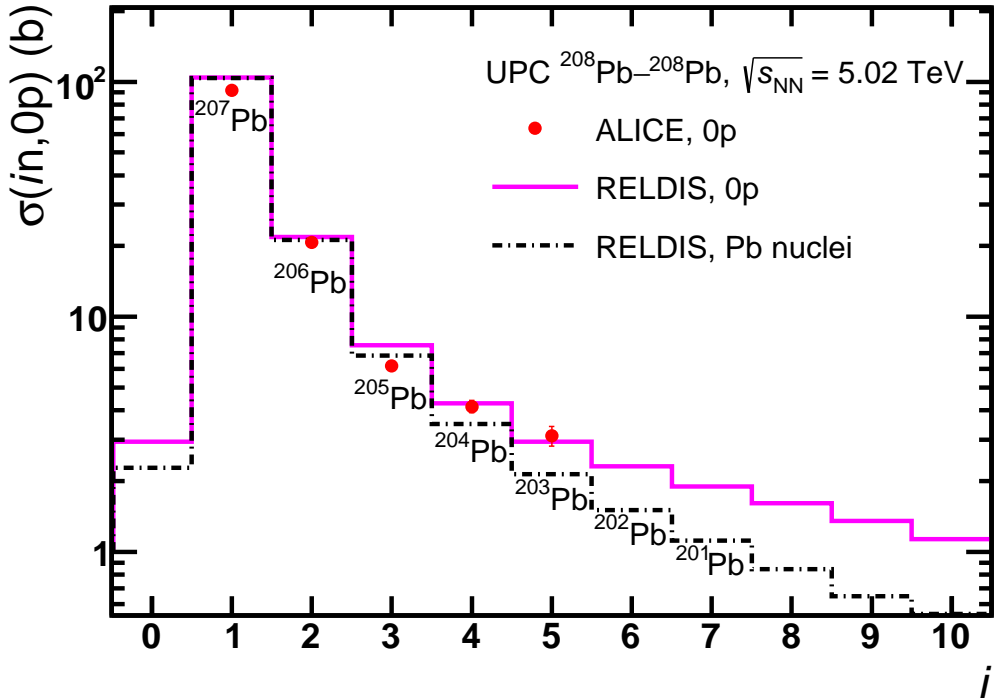}
\caption{Colour online: measured cross sections of emission of given numbers of neutrons $i$ in UPCs of $^{208}$Pb nuclei at $\sqrt{s_{\mathrm{NN}}}=5.02$~TeV without proton emission (points) and the cross sections calculated with RELDIS~\cite{Pshenichnov2011} (solid-line histogram). Calculated cross sections to produce specific secondary nuclei, $^{207,206,205,204,203,202,201}$Pb, are presented by the dash-dotted histogram marked by nuclide symbols. Combined statistical and systematic uncertainties of the measurements are presented.}  
\label{fig:neutrons0p}
\end{centering}
\end{figure}

The measured 0p neutron emission cross sections are shown in Fig.~\ref{fig:neutrons0p} and compared with RELDIS results for the same cross sections and also with the calculated cross sections to produce specific secondary nuclei: $^{207,206,205,204,203}$Pb. According to RELDIS,  the cross section of the production of $^{207}$Pb, is almost the same as (1n,0p) cross section, while the cross sections of production of $^{206}$Pb and $^{205}$Pb are smaller than (2n,0p) and (3n,0p) cross sections only by 3\% and 10\%, respectively. According to this model, the calculated 0p cross sections of emission of four, five, six, and seven neutrons can be considered as upper limits for the cross sections of production of  $^{204,203,202,201}$Pb, respectively. The difference is due to the emission of additional particles, e.g.\ protons and/or charged $\pi$ mesons, which leave residual nuclei other than Pb, as explained in Sec.~\ref{Sec:discussion}. 

\section{Discussion}\label{Sec:discussion}

The measured cross sections $\sigma(i{\mathrm{n}})$ of neutron emission accompanied by an arbitrary number of protons in EMD can be used to validate various EMD models. The cross sections of neutron emission $\sigma(i{\mathrm{n}},{\mathrm{0p}})$ without protons were also measured. Because of the absence of proton emission, these cross sections can be associated with the production of $^{207}$Pb, $^{206}$Pb, $^{205}$Pb, $^{204}$Pb, and $^{203}$Pb as secondary nuclei in EMD of $^{208}$Pb. 

\begin{figure}[t]
\begin{centering}
\includegraphics[width=1.0\columnwidth]{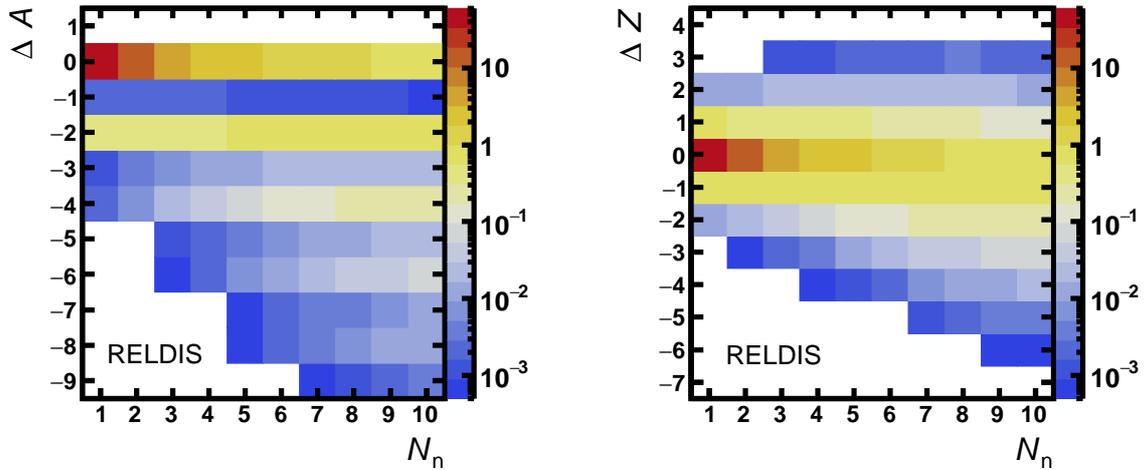}
\caption{Colour online: Probability, calculated with RELDIS (in \%), of the difference $\Delta A$ between the sum of the mass number of the most heavy residual nucleus and the numbers of emitted nucleons and $A$ of the initial $^{208}$Pb (left) and probability (in \%) of the difference $\Delta Z$ between the sum of the charge of the most heavy residual nucleus and the number of emitted protons and $Z$ of $^{208}$Pb (right) plotted as functions of  $N_{\mathrm n}$ in each event.}  
\label{fig:da_dz}
\end{centering}
\end{figure}

In order to test this association, the RELDIS event generator, validated with data on neutron emission~\cite{Golubeva2005,Abelev2012n} and production of secondary nuclei~\cite{Scheidenberger2004} in EMD of $^{208}$Pb at lower collision energies, was used. In each simulated event $\Delta A = A_{\mathrm{res}}+N_{\mathrm{n}}+N_{\mathrm{p}} - 208 $ and $\Delta Z = Z_{\mathrm{res}}+N_{\mathrm{p}} - 82$ were calculated from the mass number $A_\mathrm{res}$ and charge $Z_\mathrm{res}$ of the heaviest residual nucleus and the numbers of emitted neutrons $N_{\mathrm{n}}$ and protons $N_{\mathrm{p}}$. The calculated probability distributions to obtain $\Delta A$ and $\Delta Z$ in events with a given $N_{\mathrm{n}}$ are presented in Fig.~\ref{fig:da_dz}. As seen from this figure, in RELDIS simulations the probabilities to obtain  1n, 2n, ..., 5n events with $\Delta A=0$ or $\Delta Z=0$ among all EMD events are within the range of 3--50\%.  However, the probability of 1n, 2n, ..., 5n  events with $\Delta A<0$ or $|\Delta Z|>0$ is well below 1\%. 

Two examples of such rare events are presented in the following. The first one is represented by the excitation of an intranuclear neutron with its conversion to a proton: $\gamma {\mathrm n} \rightarrow \Delta^{\mathrm o} \rightarrow {\mathrm p} + \pi^{-}$. In the case of the undetected $\pi^{-}$ escaping the nucleus, the total charge of the system, including emitted nucleons, differs from 82: $\Delta Z=1$. In Ref.~\cite{Scheidenberger2002} the  production of $_{83}$Bi from 158A~GeV  $_{82}$Pb nuclei in their ultraperipheral collisions with different targets has been explained by the production of $\pi^{-}$. In the second example, the products of the fragmentation contain an undetected $\alpha$-particle in addition to nucleons: $\Delta A=-4$. In Ref.~\cite{Adler1974} a similar reaction of the emission of  $\alpha$-particles from $^{197}$Au induced by bremsstrahlung photons was investigated.

One can conclude from Fig.~\ref{fig:da_dz} that, according to RELDIS, in the majority of 1n--5n EMD events ${\Delta A=0}$ and ${\Delta Z= 0}$, and it is sufficient to measure $N_{\mathrm n}$ and $N_{\mathrm p}$ for evaluating $A_\mathrm{res}$ and $Z_\mathrm{res}$. The dominance of ${\Delta A= 0}$ and ${\Delta Z= 0}$ events is explained by a rather modest excitation energy per nucleon $E^\star / A$ delivered in the absorption of Weizs\"{a}cker--Williams photons by $^{208}$Pb. This was demonstrated in Ref.~\cite{Pshenichnov2005} where $\langle E^\star / A\rangle$ were calculated for photospallation of $^{\mathrm{nat}}$Pb by real photons with energies from 20~MeV to 4~GeV. As found in Ref.~\cite{Pshenichnov2005}, $\langle E^\star / A\rangle$ is typically below 1~MeV.  This suggests the creation of an excited single heavy nuclear residue as a result of the emission of free neutrons and protons during the intranuclear cascade. Such an excited nucleus would evaporate few nucleons and possibly undergo nuclear fission~\cite{Pshenichnov2005} rather than a multi-fragment break-up which becomes possible only at $E^\star / A > 3$~MeV~\cite{Bondorf1995}. For example, the decay of the giant dipole resonance in $^{208}$Pb with $E^\star =\mbox{7--20}$~MeV mostly produces one  neutron~\cite{Varlamov2021}. According to RELDIS~\cite{Pshenichnov2001}, the production of two neutrons is ten times less frequent in EMD of $^{208}$Pb in this interval of $E^\star$. The residual nuclei, $^{207}$Pb or $^{206}$Pb, can be distinguished by detecting the number of emitted neutrons. 

The contribution of electromagnetically induced fission  of $^{208}$Pb in UPCs can also be estimated with RELDIS. The fission process violates the condition $\Delta A=0$ and $\Delta Z= 0$, but its calculated probability at $\sqrt{s_{\mathrm{NN}}}=5.02$~TeV is as low as 0.6\%. The EM fission probability for 158A~GeV $^{208}$Pb projectiles on Pb target is calculated with RELDIS as 0.18\%. It is of the same order as the fission probabilities of 0.66\% and 0.75\% measured at the CERN SPS in Refs.~\cite{Alessandro2004} and~\cite{Abreu1999}, respectively.  One can conclude that both theory and experiment indicate fission probabilities below 1\%. This process can be safely neglected in favour of the creation of a single residual nucleus and several nucleons in EMD of $^{208}$Pb.

\section{Conclusions and outlook}\label{Sec:Conclusions}

The cross sections for emission of given numbers of forward neutrons in EMD of $^{208}$Pb nuclei in ultraperipheral collisions at $\sqrt{s_{\mathrm{NN}}}=5.02$~TeV have been measured with the ALICE neutron zero degree calorimeters. The fractions of 1n, 2n, and 3n events were measured at the highest $^{208}$Pb--$^{208}$Pb collision energy available so far in accelerator experiments. These fractions were found to be similar to those measured at $\sqrt{s_{\mathrm{NN}}}=2.76$~TeV. The predictions from the  RELDIS~\cite{Pshenichnov2011} and $\mathrm{n^O_On}$~\cite{Broz2020} models describe well the measured cross sections, in particular, for low neutron multiplicities. These measurements are important for the extraction of the contributions of high- and low-energy photons from coherent vector meson photoproduction measurements accompanied by neutron emission~\cite{Guzey2013,Citron2018}.

The cross sections of EMD events with the emission of exactly  one, two, three, four, and five neutrons and without emission of protons have been measured. According to the RELDIS model, in EMD a single heavy residual nucleus is typically produced after the emission of several nucleons. The probability of nuclear fission in EMD is estimated to be below 1\%. Therefore, the EMD events of 1n, 2n, 3n, 4n, and 5n emission without protons can be associated with the production of $^{207}$Pb, $^{206}$Pb, $^{205}$Pb, $^{204}$Pb, and $^{203}$Pb, respectively. In other words, the measured neutron emission cross sections can be considered as upper limits for the respective isotope production cross sections.  

Since the charge-to-mass ratio of $^{207}$Pb remains close to $^{208}$Pb, these frequently produced secondary nuclei emerge from the interaction points and propagate in the field of the LHC  magnets on dispersive trajectories in the vicinity of the primary beam~\cite{Bruce2009}. The collider collimation system is adjusted to intercept them (off-momentum collimators around Point 3 of LHC) to avoid the risk of beam dumps or quenches of superconducting magnets~\cite{Klein2001,Bruce2009,Hermes2016,Schaumann2020}. The moderate fluxes of $^{206}$Pb nuclei may hit the beam pipe closer to the interaction points. The smaller fluxes of  $^{205,204,\ldots}$Pb generated in the collisions are lost at the start of the dispersion suppressor or taken up by physics debris absorbers and do not present any risks in collider operation.

A good quantitative understanding of the cross sections and  fluxes of these nuclei  provides valuable input for evaluating luminosity decay and beam losses in the design of the Future Circular Collider (FCC-hh)~\cite{Schaumann2015}.  They also increase confidence in similar calculations for interactions of beam nuclei with carbon and other materials in the collimation systems of the LHC and FCC~\cite{Hermes2016}.


\newenvironment{acknowledgement}{\relax}{\relax}
\begin{acknowledgement}
\section*{Acknowledgements}

The ALICE Collaboration would like to thank all its engineers and technicians for their invaluable contributions to the construction of the experiment and the CERN accelerator teams for the outstanding performance of the LHC complex.
The ALICE Collaboration gratefully acknowledges the resources and support provided by all Grid centres and the Worldwide LHC Computing Grid (WLCG) collaboration.
The ALICE Collaboration acknowledges the following funding agencies for their support in building and running the ALICE detector:
A. I. Alikhanyan National Science Laboratory (Yerevan Physics Institute) Foundation (ANSL), State Committee of Science and World Federation of Scientists (WFS), Armenia;
Austrian Academy of Sciences, Austrian Science Fund (FWF): [M 2467-N36] and Nationalstiftung f\"{u}r Forschung, Technologie und Entwicklung, Austria;
Ministry of Communications and High Technologies, National Nuclear Research Center, Azerbaijan;
Conselho Nacional de Desenvolvimento Cient\'{\i}fico e Tecnol\'{o}gico (CNPq), Financiadora de Estudos e Projetos (Finep), Funda\c{c}\~{a}o de Amparo \`{a} Pesquisa do Estado de S\~{a}o Paulo (FAPESP) and Universidade Federal do Rio Grande do Sul (UFRGS), Brazil;
Bulgarian Ministry of Education and Science, within the National Roadmap for Research Infrastructures 2020¿2027 (object CERN), Bulgaria;
Ministry of Education of China (MOEC) , Ministry of Science \& Technology of China (MSTC) and National Natural Science Foundation of China (NSFC), China;
Ministry of Science and Education and Croatian Science Foundation, Croatia;
Centro de Aplicaciones Tecnol\'{o}gicas y Desarrollo Nuclear (CEADEN), Cubaenerg\'{\i}a, Cuba;
Ministry of Education, Youth and Sports of the Czech Republic, Czech Republic;
The Danish Council for Independent Research | Natural Sciences, the VILLUM FONDEN and Danish National Research Foundation (DNRF), Denmark;
Helsinki Institute of Physics (HIP), Finland;
Commissariat \`{a} l'Energie Atomique (CEA) and Institut National de Physique Nucl\'{e}aire et de Physique des Particules (IN2P3) and Centre National de la Recherche Scientifique (CNRS), France;
Bundesministerium f\"{u}r Bildung und Forschung (BMBF) and GSI Helmholtzzentrum f\"{u}r Schwerionenforschung GmbH, Germany;
General Secretariat for Research and Technology, Ministry of Education, Research and Religions, Greece;
National Research, Development and Innovation Office, Hungary;
Department of Atomic Energy Government of India (DAE), Department of Science and Technology, Government of India (DST), University Grants Commission, Government of India (UGC) and Council of Scientific and Industrial Research (CSIR), India;
National Research and Innovation Agency - BRIN, Indonesia;
Istituto Nazionale di Fisica Nucleare (INFN), Italy;
Japanese Ministry of Education, Culture, Sports, Science and Technology (MEXT) and Japan Society for the Promotion of Science (JSPS) KAKENHI, Japan;
Consejo Nacional de Ciencia (CONACYT) y Tecnolog\'{i}a, through Fondo de Cooperaci\'{o}n Internacional en Ciencia y Tecnolog\'{i}a (FONCICYT) and Direcci\'{o}n General de Asuntos del Personal Academico (DGAPA), Mexico;
Nederlandse Organisatie voor Wetenschappelijk Onderzoek (NWO), Netherlands;
The Research Council of Norway, Norway;
Commission on Science and Technology for Sustainable Development in the South (COMSATS), Pakistan;
Pontificia Universidad Cat\'{o}lica del Per\'{u}, Peru;
Ministry of Education and Science, National Science Centre and WUT ID-UB, Poland;
Korea Institute of Science and Technology Information and National Research Foundation of Korea (NRF), Republic of Korea;
Ministry of Education and Scientific Research, Institute of Atomic Physics, Ministry of Research and Innovation and Institute of Atomic Physics and University Politehnica of Bucharest, Romania;
Ministry of Education, Science, Research and Sport of the Slovak Republic, Slovakia;
National Research Foundation of South Africa, South Africa;
Swedish Research Council (VR) and Knut \& Alice Wallenberg Foundation (KAW), Sweden;
European Organization for Nuclear Research, Switzerland;
Suranaree University of Technology (SUT), National Science and Technology Development Agency (NSTDA), Thailand Science Research and Innovation (TSRI) and National Science, Research and Innovation Fund (NSRF), Thailand;
Turkish Energy, Nuclear and Mineral Research Agency (TENMAK), Turkey;
National Academy of  Sciences of Ukraine, Ukraine;
Science and Technology Facilities Council (STFC), United Kingdom;
National Science Foundation of the United States of America (NSF) and United States Department of Energy, Office of Nuclear Physics (DOE NP), United States of America.
In addition, individual groups or members have received support from:
Marie Sk\l{}odowska Curie, European Research Council, Strong 2020 - Horizon 2020 (grant nos. 950692, 824093, 896850), European Union;
Academy of Finland (Center of Excellence in Quark Matter) (grant nos. 346327, 346328), Finland;
Programa de Apoyos para la Superaci\'{o}n del Personal Acad\'{e}mico, UNAM, Mexico.

\end{acknowledgement}

\bibliographystyle{utphys}   
\bibliography{NeutronsPbPaper}

\providecommand{\href}[2]{#2}\begingroup\raggedright\begin{thebibliography}{10}

\bibitem{Roland2014}
G.~Roland, K.~Safarik, and P.~Steinberg, ``{Heavy-ion collisions at the LHC}'',
  \href{http://dx.doi.org/10.1016/j.ppnp.2014.05.001}{{\em Prog. Part. Nucl.
  Phys.} {\bfseries 77} (2014) 70--127}.

\bibitem{Appelshauser1998}
{\bfseries NA49} Collaboration, H.~Appelsh\"auser {\em et~al.}, ``{Spectator
  nucleons in Pb + Pb collisions at 158-A-GeV}'',
  \href{http://dx.doi.org/10.1007/s100500050135}{{\em Eur. Phys. J. A}
  {\bfseries 2} (1998) 383--390}.

\bibitem{Scheidenberger2004}
C.~Scheidenberger {\em et~al.}, ``{Charge-changing interactions of
  ultrarelativistic Pb nuclei}'',
  \href{http://dx.doi.org/10.1103/PhysRevC.70.014902}{{\em Phys. Rev. C}
  {\bfseries 70} (2004) 014902}.

\bibitem{Pshenichnov2011}
I.~A. Pshenichnov, ``{Electromagnetic excitation and fragmentation of
  ultrarelativistic nuclei}'',
  \href{http://dx.doi.org/10.1134/S1063779611020067}{{\em Phys. Part. Nucl.}
  {\bfseries 42} (2011) 215--250}.

\bibitem{Abelev2012n}
{\bfseries ALICE} Collaboration, B.~Abelev {\em et~al.}, ``{Measurement of the
  cross section for electromagnetic dissociation with neutron emission in
  Pb--Pb collisions at $\sqrt{s_{NN}}$ = 2.76 TeV}'',
  \href{http://dx.doi.org/10.1103/PhysRevLett.109.252302}{{\em Phys. Rev.
  Lett.} {\bfseries 109} (2012) 252302},
  \href{http://arxiv.org/abs/1203.2436}{{\ttfamily arXiv:1203.2436 [nucl-ex]}}.

\bibitem{Baltz1996}
A.~J. Baltz, M.~J. Rhoades-Brown, and J.~Weneser, ``{Heavy ion partial beam
  lifetimes due to Coulomb induced processes}'',
  \href{http://dx.doi.org/10.1103/PhysRevE.54.4233}{{\em Phys. Rev. E}
  {\bfseries 54} (1996) 4233--4239}.

\bibitem{Bruce2009}
R.~Bruce, D.~Bocian, S.~Gilardoni, and J.~M. Jowett, ``{Beam losses from
  ultraperipheral nuclear collisions between $^{208}$Pb$^{82+}$ ions in the
  Large Hadron Collider and their alleviation}'',
  \href{http://dx.doi.org/10.1103/PhysRevSTAB.12.071002}{{\em Phys. Rev. ST
  Accel. Beams} {\bfseries 12} (2009) 071002}.

\bibitem{Hermes2016}
P.~D. Hermes, R.~Bruce, J.~M. Jowett, S.~Redaelli, B.~Salvachua~Ferrando,
  G.~Valentino, and D.~Wollmann, ``{Measured and simulated heavy-ion beam loss
  patterns at the CERN Large Hadron Collider}'',
  \href{http://dx.doi.org/10.1016/j.nima.2016.02.050}{{\em Nucl. Instrum. Meth.
  A} {\bfseries 819} (2016) 73--83}.

\bibitem{Botvina1995}
A.~S. Botvina {\em et~al.}, ``{Multifragmentation of spectators in relativistic
  heavy ion reactions}'',
  \href{http://dx.doi.org/10.1016/0375-9474(94)00621-S}{{\em Nucl. Phys. A}
  {\bfseries 584} (1995) 737--756}.

\bibitem{Schuttauf1996}
A.~Sch\"uttauf {\em et~al.}, ``{Universality of spectator fragmentation at
  relativistic bombarding energies}'',
  \href{http://dx.doi.org/10.1016/0375-9474(96)00239-4}{{\em Nucl. Phys. A}
  {\bfseries 607} (1996) 457--486},
  \href{http://arxiv.org/abs/nucl-ex/9606001}{{\ttfamily
  arXiv:nucl-ex/9606001}}.

\bibitem{Deines-Jones2000}
P.~Deines-Jones {\em et~al.}, ``{Charged particle production in the Pb + Pb
  system at 158~GeV/c per nucleon}'',
  \href{http://dx.doi.org/10.1103/PhysRevC.62.014903}{{\em Phys. Rev. C}
  {\bfseries 62} (2000) 014903},
  \href{http://arxiv.org/abs/hep-ex/9912008}{{\ttfamily arXiv:hep-ex/9912008}}.

\bibitem{Cecchini2002}
S.~Cecchini, G.~Giacomelli, M.~Giorgini, G.~Mandrioli, L.~Patrizii, V.~Popa,
  P.~Serra, G.~Sirri, and M.~Spurio, ``{Fragmentation cross-sections of
  158A~GeV Pb ions in various targets measured with CR39 nuclear track
  detectors}'', \href{http://dx.doi.org/10.1016/S0375-9474(02)00962-4}{{\em
  Nucl. Phys. A} {\bfseries 707} (2002) 513--524},
  \href{http://arxiv.org/abs/hep-ex/0201039}{{\ttfamily arXiv:hep-ex/0201039}}.

\bibitem{Uggerhjoj:In}
U.~Uggerh{\o}j, I.~A. Pshenichnov, C.~Scheidenberger, H.~D. Hansen, H.~Knudsen,
  E.~Uggerh{\o}j, P.~Sona, A.~Mangiarotti, and S.~Ballestrero,
  ``{Charge-changing interactions of ultrarelativistic In nuclei}'',
  \href{http://dx.doi.org/10.1103/PhysRevC.72.057901}{{\em Phys. Rev. C}
  {\bfseries 72} (2005) 057901}.

\bibitem{Tarafdar2014}
S.~Tarafdar, Z.~Citron, and A.~Milov, ``{A Centrality Detector Concept}'',
  \href{http://dx.doi.org/10.1016/j.nima.2014.09.060}{{\em Nucl. Instrum. Meth.
  A} {\bfseries 768} (2014) 170--178},
  \href{http://arxiv.org/abs/1405.4555}{{\ttfamily arXiv:1405.4555 [nucl-ex]}}.

\bibitem{Grinstein2016}
{\bfseries AFP} Collaboration, S.~Grinstein, ``{The ATLAS Forward Proton
  Detector (AFP)}'',
  \href{http://dx.doi.org/10.1016/j.nuclphysbps.2015.09.185}{{\em Nucl. Part.
  Phys. Proc.} {\bfseries 273-275} (2016) 1180--1184}.

\bibitem{Puddu2007}
G.~Puddu {\em et~al.}, ``{The zero degree calorimeters for the ALICE
  experiment}'', \href{http://dx.doi.org/10.1016/j.nima.2007.08.013}{{\em Nucl.
  Instrum. Meth. Phys. Res. Sect. A} {\bfseries 581} (2007) 397--401}.

\bibitem{Gemme2009}
R.~Gemme {\em et~al.}, ``{Commissioning and calibration of the zero degree
  calorimeters for the ALICE experiment}'',
  \href{http://dx.doi.org/10.1016/j.nuclphysbps.2009.10.069}{{\em Nucl. Phys. B
  Proc. Suppl.} {\bfseries 197} (2009) 211--214}.

\bibitem{Oppedisano2009}
C.~Oppedisano {\em et~al.}, ``{Physics performance of the ALICE zero degree
  calorimeter}'',
  \href{http://dx.doi.org/10.1016/j.nuclphysbps.2009.10.068}{{\em Nucl. Phys. B
  Proc. Suppl.} {\bfseries 197} (2009) 206--210}.

\bibitem{Pshenichnov1999}
I.~A. Pshenichnov, I.~N. Mishustin, J.~P. Bondorf, A.~S. Botvina, and A.~S.
  Ilinov, ``{Particle emission following Coulomb excitation in
  ultrarelativistic heavy ion collisions}'',
  \href{http://dx.doi.org/10.1103/PhysRevC.60.044901}{{\em Phys. Rev. C}
  {\bfseries 60} (1999) 044901},
  \href{http://arxiv.org/abs/nucl-th/9901061}{{\ttfamily
  arXiv:nucl-th/9901061}}.

\bibitem{Pshenichnov2001}
I.~A. Pshenichnov, J.~P. Bondorf, I.~N. Mishustin, A.~Ventura, and S.~Masetti,
  ``{Mutual heavy ion dissociation in peripheral collisions at
  ultrarelativistic energies}'',
  \href{http://dx.doi.org/10.1103/PhysRevC.64.024903}{{\em Phys. Rev. C}
  {\bfseries 64} (2001) 024903},
  \href{http://arxiv.org/abs/nucl-th/0101035}{{\ttfamily
  arXiv:nucl-th/0101035}}.

\bibitem{Braun2014}
H.~H. Braun, A.~Fass\`o, A.~Ferrari, J.~M. Jowett, P.~R. Sala, and G.~I.
  Smirnov, ``{Hadronic and electromagnetic fragmentation of ultrarelativistic
  heavy ions at LHC}'',
  \href{http://dx.doi.org/10.1103/PhysRevSTAB.17.021006}{{\em Phys. Rev. ST
  Accel. Beams} {\bfseries 17} (2014) 021006}.

\bibitem{Klusek-Gawenda2014}
M.~K\l{}usek-Gawenda, M.~Ciema\l{}a, W.~Sch{\"a}fer, and A.~Szczurek,
  ``{Electromagnetic excitation of nuclei and neutron evaporation in
  ultrarelativistic ultraperipheral heavy ion collisions}'',
  \href{http://dx.doi.org/10.1103/PhysRevC.89.054907}{{\em Phys. Rev. C}
  {\bfseries 89} (2014) 054907},
  \href{http://arxiv.org/abs/1311.1938}{{\ttfamily arXiv:1311.1938 [nucl-th]}}.

\bibitem{Broz2020}
M.~Broz, J.~G. Contreras, and J.~D. Tapia~Takaki, ``{{A generator of forward
  neutrons for ultra-peripheral collisions: $\mathrm{n^O_On}$}}'',
  \href{http://dx.doi.org/10.1016/j.cpc.2020.107181}{{\em Comput. Phys.
  Commun.} {\bfseries 253} (2020) 107181},
  \href{http://arxiv.org/abs/1908.08263}{{\ttfamily arXiv:1908.08263
  [nucl-th]}}.

\bibitem{ALICE2008JINST}
{\bfseries ALICE} Collaboration, K.~Aamodt {\em et~al.}, ``{The ALICE
  experiment at the CERN LHC}'',
  \href{http://dx.doi.org/10.1088/1748-0221/3/08/S08002}{{\em JINST} {\bfseries
  3} (2008) S08002}.

\bibitem{vanderMeer1968}
S.~van~der Meer, ``{Calibration of the effective beam height in the ISR}'',
  tech. rep., CERN, Geneva, 1968.
\newblock \url{https://cds.cern.ch/record/296752}.

\bibitem{Castellanos2021}
{\bfseries ALICE} Collaboration, S.~Acharya, D.~Adamov{\'a}, A.~Adler,
  J.~Adolfsson, {\em et~al.}, ``{ALICE luminosity determination for Pb$-$Pb
  collisions at $\sqrt{s_{\mathrm{NN}}} = 5.02$ TeV}'',
  \href{http://arxiv.org/abs/2204.10148}{{\ttfamily arXiv:2204.10148
  [nucl-ex]}}.

\bibitem{Svetlichnyi2020}
A.~O. Svetlichnyi and I.~A. Pshenichnov, ``{Formation of Free and Bound
  Spectator Nucleons in Hadronic Interactions between Relativistic Nuclei}'',
  \href{http://dx.doi.org/10.3103/S1062873820080110}{{\em Bull. Russ. Acad.
  Sci. Phys.} {\bfseries 84} (2020) 911--916}.

\bibitem{Casadei:2009ic}
D.~Casadei, ``{Estimating the selection efficiency}'',
  \href{http://dx.doi.org/10.1088/1748-0221/7/08/P08021}{{\em JINST} {\bfseries
  7} (2012) P08021}, \href{http://arxiv.org/abs/0908.0130}{{\ttfamily
  arXiv:0908.0130 [physics.data-an]}}.

\bibitem{Baltz2009}
A.~J. Baltz, Y.~Gorbunov, S.~R. Klein, and J.~Nystrand, ``{Two-photon
  interactions with nuclear breakup in relativistic heavy ion collisions}'',
  \href{http://dx.doi.org/10.1103/PhysRevC.80.044902}{{\em Phys. Rev. C}
  {\bfseries 80} (2009) 044902},
  \href{http://arxiv.org/abs/0907.1214}{{\ttfamily arXiv:0907.1214 [nucl-ex]}}.

\bibitem{Klein2017}
S.~R. Klein, J.~Nystrand, J.~Seger, Y.~Gorbunov, and J.~Butterworth,
  ``{{STARlight: A Monte Carlo simulation program for ultra-peripheral
  collisions of relativistic ions}}'',
  \href{http://dx.doi.org/10.1016/j.cpc.2016.10.016}{{\em Comput. Phys.
  Commun.} {\bfseries 212} (2017) 258--268},
  \href{http://arxiv.org/abs/1607.03838}{{\ttfamily arXiv:1607.03838
  [hep-ph]}}.

\bibitem{Dmitrieva2018}
U.~Dmitrieva and I.~Pshenichnov, ``{On the performance of Zero Degree
  Calorimeters in detecting multinucleon events}'',
  \href{http://dx.doi.org/10.1016/j.nima.2018.07.072}{{\em Nucl. Instrum. Meth.
  A} {\bfseries 906} (2018) 114--119},
  \href{http://arxiv.org/abs/1805.01792}{{\ttfamily arXiv:1805.01792
  [physics.ins-det]}}.

\bibitem{Golubeva2005}
M.~B. Golubeva {\em et~al.}, ``{Neutron emission in electromagnetic
  dissociation of ultrarelativistic Pb ions}'',
  \href{http://dx.doi.org/10.1103/PhysRevC.71.024905}{{\em Phys. Rev. C}
  {\bfseries 71} (2005) 024905}.

\bibitem{Pshenichnov2005}
I.~A. Pshenichnov, B.~L. Berman, W.~J. Briscoe, C.~Cetina, G.~Feldman,
  P.~Heimberg, A.~S. Iljinov, and I.~I. Strakovsky, ``{Intranuclear-cascade
  model calculation of photofission probabilities for actinide nuclei}'',
  \href{http://dx.doi.org/10.1140/epja/i2004-10130-9}{{\em Eur. Phys. J. A}
  {\bfseries 24} (2005) 69--84},
  \href{http://arxiv.org/abs/nucl-th/0303070}{{\ttfamily
  arXiv:nucl-th/0303070}}.

\bibitem{Pshenichnov2019}
I.~A. Pshenichnov and S.~A. Gunin, ``{Electromagnetic interactions of nuclei at
  the FCC-hh collider}'',
  \href{http://dx.doi.org/10.1134/S1063779619050198}{{\em Phys. Part. Nucl.}
  {\bfseries 50} (2019) 501--505}.

\bibitem{Muccifora1999}
V.~Muccifora {\em et~al.}, ``{Photoabsorption on nuclei in the shadowing
  threshold region}'', \href{http://dx.doi.org/10.1103/PhysRevC.60.064616}{{\em
  Phys. Rev. C} {\bfseries 60} (1999) 064616},
  \href{http://arxiv.org/abs/nucl-ex/9810015}{{\ttfamily
  arXiv:nucl-ex/9810015}}.

\bibitem{Scheidenberger2002}
C.~Scheidenberger {\em et~al.}, ``{Electromagnetically induced nuclear-charge
  pickup observed in ultrarelativistic Pb collisions}'',
  \href{http://dx.doi.org/10.1103/PhysRevLett.88.042301}{{\em Phys. Rev. Lett.}
  {\bfseries 88} (2002) 042301}.

\bibitem{Adler1974}
J.~O. Adler, G.~Andersson, and H.~A. Gustafsson, ``{{Alpha particles from
  $^{197}$Au irradiated by 500~MeV bremsstrahlung}}'',
  \href{http://dx.doi.org/10.1016/0375-9474(74)90282-6}{{\em Nucl. Phys. A}
  {\bfseries 223} (1974) 145--156}.

\bibitem{Bondorf1995}
J.~P. Bondorf, A.~S. Botvina, A.~S. Ilinov, I.~N. Mishustin, and K.~Sneppen,
  ``{Statistical multifragmentation of nuclei}'',
  \href{http://dx.doi.org/10.1016/0370-1573(94)00097-M}{{\em Phys. Rept.}
  {\bfseries 257} (1995) 133--221}.

\bibitem{Varlamov2021}
V.~V. Varlamov, A.~I. Davydov, and V.~N. Orlin, ``{New evaluated data on
  $^{206,207,208}$Pb photodisintegration}'',
  \href{http://dx.doi.org/10.1140/epja/s10050-021-00603-8}{{\em Eur. Phys. J.
  A} {\bfseries 57} (2021) 287}.

\bibitem{Alessandro2004}
B.~Alessandro {\em et~al.}, ``{Fission cross sections of lead projectiles in Pb
  nucleus interactions at 40 and 158~GeV/c per nucleon}'',
  \href{http://dx.doi.org/10.1103/PhysRevC.69.034904}{{\em Phys. Rev. C}
  {\bfseries 69} (2004) 034904}.

\bibitem{Abreu1999}
{\bfseries NA50} Collaboration, M.~C. Abreu {\em et~al.}, ``{Observation of
  fission in Pb--Pb interactions at 158A~GeV}'',
  \href{http://dx.doi.org/10.1103/PhysRevC.59.876}{{\em Phys. Rev. C}
  {\bfseries 59} (1999) 876--883}.

\bibitem{Guzey2013}
V.~Guzey, M.~Strikman, and M.~Zhalov, ``{Disentangling coherent and incoherent
  quasielastic $J/\psi$ photoproduction on nuclei by neutron tagging in
  ultraperipheral ion collisions at the LHC}'',
  \href{http://dx.doi.org/10.1140/epjc/s10052-014-2942-z}{{\em Eur. Phys. J. C}
  {\bfseries 74} (2014) 2942}, \href{http://arxiv.org/abs/1312.6486}{{\ttfamily
  arXiv:1312.6486 [hep-ph]}}.

\bibitem{Citron2018}
Z.~Citron {\em et~al.}, ``{Report from Working Group 5}: {Future physics
  opportunities for high-density QCD at the LHC with heavy-ion and proton
  beams}'', \href{http://dx.doi.org/10.23731/CYRM-2019-007.1159}{{\em CERN
  Yellow Rep. Monogr.} {\bfseries 7} (2019) 1159--1410},
  \href{http://arxiv.org/abs/1812.06772}{{\ttfamily arXiv:1812.06772
  [hep-ph]}}.

\bibitem{Klein2001}
S.~R. Klein, ``{Localized beam pipe heating due to $\mathrm{e}^-$ capture and
  nuclear excitation in heavy ion colliders}'',
  \href{http://dx.doi.org/10.1016/S0168-9002(00)00995-5}{{\em Nucl. Instrum.
  Meth. A} {\bfseries 459} (2001) 51--57},
  \href{http://arxiv.org/abs/physics/0005032}{{\ttfamily
  arXiv:physics/0005032}}.

\bibitem{Schaumann2020}
M.~Schaumann, J.~M. Jowett, C.~Bahamonde~Castro, R.~Bruce, A.~Lechner, and
  T.~Mertens, ``{Bound-free pair production from nuclear collisions and the
  steady-state quench limit of the main dipole magnets of the CERN Large Hadron
  Collider}'',
  \href{http://dx.doi.org/10.1103/PhysRevAccelBeams.23.121003}{{\em Phys. Rev.
  Accel. Beams} {\bfseries 23} (2020) 121003},
  \href{http://arxiv.org/abs/2008.05312}{{\ttfamily arXiv:2008.05312
  [physics.acc-ph]}}.

\bibitem{Schaumann2015}
M.~Schaumann, ``{Potential performance for Pb-Pb, p-Pb and p-p collisions in a
  future circular collider}'',
  \href{http://dx.doi.org/10.1103/PhysRevSTAB.18.091002}{{\em Phys. Rev. ST
  Accel. Beams} {\bfseries 18} (2015) 091002},
  \href{http://arxiv.org/abs/1503.09107}{{\ttfamily arXiv:1503.09107
  [physics.acc-ph]}}.

\end{thebibliography}\endgroup

\newpage
\appendix

%
%

\section{The ALICE Collaboration}
\label{app:collab}
\begin{flushleft} 
\small

S.~Acharya\,\orcidlink{0000-0002-9213-5329}\,$^{\rm 124}$, 
D.~Adamov\'{a}\,\orcidlink{0000-0002-0504-7428}\,$^{\rm 86}$, 
A.~Adler$^{\rm 69}$, 
G.~Aglieri Rinella\,\orcidlink{0000-0002-9611-3696}\,$^{\rm 32}$, 
M.~Agnello\,\orcidlink{0000-0002-0760-5075}\,$^{\rm 29}$, 
N.~Agrawal\,\orcidlink{0000-0003-0348-9836}\,$^{\rm 50}$, 
Z.~Ahammed\,\orcidlink{0000-0001-5241-7412}\,$^{\rm 131}$, 
S.~Ahmad\,\orcidlink{0000-0003-0497-5705}\,$^{\rm 15}$, 
S.U.~Ahn\,\orcidlink{0000-0001-8847-489X}\,$^{\rm 70}$, 
I.~Ahuja\,\orcidlink{0000-0002-4417-1392}\,$^{\rm 37}$
% ...

\section*{Affiliation Notes}

$^{\rm I}$ Deceased\\
$^{\rm II}$ Also at: Max-Planck-Institut f\"{u}r Physik, Munich, Germany\\
$^{\rm III}$ Also at: Italian National Agency for New Technologies, Energy and Sustainable Economic Development (ENEA), Bologna, Italy\\
$^{\rm IV}$ Also at: Dipartimento DET del Politecnico di Torino, Turin, Italy\\
$^{\rm V}$ Also at: Department of Applied Physics, Aligarh Muslim University, Aligarh, India\\
$^{\rm VI}$ Also at: Institute of Theoretical Physics, University of Wroclaw, Poland\\
$^{\rm VII}$ Also at: An institution covered by a cooperation agreement with CERN\\

\section*{Collaboration Institutes}

$^{1}$ A.I. Alikhanyan National Science Laboratory (Yerevan Physics Institute) Foundation, Yerevan, Armenia\\
$^{2}$ AGH University of Science and Technology, Cracow, Poland\\
$^{3}$ Bogolyubov Institute for Theoretical Physics, National Academy of Sciences of Ukraine, Kiev, Ukraine\\
$^{4}$ Bose Institute, Department of Physics  and Centre for Astroparticle Physics and Space Science (CAPSS), Kolkata, India\\
$^{5}$ California Polytechnic State University, San Luis Obispo, California, United States\\
$^{6}$ Central China Normal University, Wuhan, China\\
$^{7}$ Centro de Aplicaciones Tecnol\'{o}gicas y Desarrollo Nuclear (CEADEN), Havana, Cuba\\
$^{8}$ Centro de Investigaci\'{o}n y de Estudios Avanzados (CINVESTAV), Mexico City and M\'{e}rida, Mexico\\
$^{9}$ Chicago State University, Chicago, Illinois, United States\\
$^{10}$ China Institute of Atomic Energy, Beijing, China\\
$^{11}$ Chungbuk National University, Cheongju, Republic of Korea\\
$^{12}$ Comenius University Bratislava, Faculty of Mathematics, Physics and Informatics, Bratislava, Slovak Republic\\
$^{13}$ COMSATS University Islamabad, Islamabad, Pakistan\\
$^{14}$ Creighton University, Omaha, Nebraska, United States\\
$^{15}$ Department of Physics, Aligarh Muslim University, Aligarh, India\\
$^{16}$ Department of Physics, Pusan National University, Pusan, Republic of Korea\\
$^{17}$ Department of Physics, Sejong University, Seoul, Republic of Korea\\
$^{18}$ Department of Physics, University of California, Berkeley, California, United States\\
$^{19}$ Department of Physics, University of Oslo, Oslo, Norway\\
$^{20}$ Department of Physics and Technology, University of Bergen, Bergen, Norway\\
$^{21}$ Dipartimento di Fisica, Universit\`{a} di Pavia, Pavia, Italy\\
$^{22}$ Dipartimento di Fisica dell'Universit\`{a} and Sezione INFN, Cagliari, Italy\\
$^{23}$ Dipartimento di Fisica dell'Universit\`{a} and Sezione INFN, Trieste, Italy\\
$^{24}$ Dipartimento di Fisica dell'Universit\`{a} and Sezione INFN, Turin, Italy\\
$^{25}$ Dipartimento di Fisica e Astronomia dell'Universit\`{a} and Sezione INFN, Bologna, Italy\\
$^{26}$ Dipartimento di Fisica e Astronomia dell'Universit\`{a} and Sezione INFN, Catania, Italy\\
$^{27}$ Dipartimento di Fisica e Astronomia dell'Universit\`{a} and Sezione INFN, Padova, Italy\\
$^{28}$ Dipartimento di Fisica `E.R.~Caianiello' dell'Universit\`{a} and Gruppo Collegato INFN, Salerno, Italy\\
$^{29}$ Dipartimento DISAT del Politecnico and Sezione INFN, Turin, Italy\\
$^{30}$ Dipartimento di Scienze MIFT, Universit\`{a} di Messina, Messina, Italy\\
$^{31}$ Dipartimento Interateneo di Fisica `M.~Merlin' and Sezione INFN, Bari, Italy\\
$^{32}$ European Organization for Nuclear Research (CERN), Geneva, Switzerland\\
$^{33}$ Faculty of Electrical Engineering, Mechanical Engineering and Naval Architecture, University of Split, Split, Croatia\\
$^{34}$ Faculty of Engineering and Science, Western Norway University of Applied Sciences, Bergen, Norway\\
$^{35}$ Faculty of Nuclear Sciences and Physical Engineering, Czech Technical University in Prague, Prague, Czech Republic\\
$^{36}$ Faculty of Physics, Sofia University, Sofia, Bulgaria\\
$^{37}$ Faculty of Science, P.J.~\v{S}af\'{a}rik University, Ko\v{s}ice, Slovak Republic\\
$^{38}$ Frankfurt Institute for Advanced Studies, Johann Wolfgang Goethe-Universit\"{a}t Frankfurt, Frankfurt, Germany\\
$^{39}$ Fudan University, Shanghai, China\\
$^{40}$ Gangneung-Wonju National University, Gangneung, Republic of Korea\\
$^{41}$ Gauhati University, Department of Physics, Guwahati, India\\
$^{42}$ Helmholtz-Institut f\"{u}r Strahlen- und Kernphysik, Rheinische Friedrich-Wilhelms-Universit\"{a}t Bonn, Bonn, Germany\\
$^{43}$ Helsinki Institute of Physics (HIP), Helsinki, Finland\\
$^{44}$ High Energy Physics Group,  Universidad Aut\'{o}noma de Puebla, Puebla, Mexico\\
$^{45}$ Horia Hulubei National Institute of Physics and Nuclear Engineering, Bucharest, Romania\\
$^{46}$ Indian Institute of Technology Bombay (IIT), Mumbai, India\\
$^{47}$ Indian Institute of Technology Indore, Indore, India\\
$^{48}$ INFN, Laboratori Nazionali di Frascati, Frascati, Italy\\
$^{49}$ INFN, Sezione di Bari, Bari, Italy\\
$^{50}$ INFN, Sezione di Bologna, Bologna, Italy\\
$^{51}$ INFN, Sezione di Cagliari, Cagliari, Italy\\
$^{52}$ INFN, Sezione di Catania, Catania, Italy\\
$^{53}$ INFN, Sezione di Padova, Padova, Italy\\
$^{54}$ INFN, Sezione di Pavia, Pavia, Italy\\
$^{55}$ INFN, Sezione di Torino, Turin, Italy\\
$^{56}$ INFN, Sezione di Trieste, Trieste, Italy\\
$^{57}$ Inha University, Incheon, Republic of Korea\\
$^{58}$ Institute for Gravitational and Subatomic Physics (GRASP), Utrecht University/Nikhef, Utrecht, Netherlands\\
$^{59}$ Institute of Experimental Physics, Slovak Academy of Sciences, Ko\v{s}ice, Slovak Republic\\
$^{60}$ Institute of Physics, Homi Bhabha National Institute, Bhubaneswar, India\\
$^{61}$ Institute of Physics of the Czech Academy of Sciences, Prague, Czech Republic\\
$^{62}$ Institute of Space Science (ISS), Bucharest, Romania\\
$^{63}$ Institut f\"{u}r Kernphysik, Johann Wolfgang Goethe-Universit\"{a}t Frankfurt, Frankfurt, Germany\\
$^{64}$ Instituto de Ciencias Nucleares, Universidad Nacional Aut\'{o}noma de M\'{e}xico, Mexico City, Mexico\\
$^{65}$ Instituto de F\'{i}sica, Universidade Federal do Rio Grande do Sul (UFRGS), Porto Alegre, Brazil\\
$^{66}$ Instituto de F\'{\i}sica, Universidad Nacional Aut\'{o}noma de M\'{e}xico, Mexico City, Mexico\\
$^{67}$ iThemba LABS, National Research Foundation, Somerset West, South Africa\\
$^{68}$ Jeonbuk National University, Jeonju, Republic of Korea\\
$^{69}$ Johann-Wolfgang-Goethe Universit\"{a}t Frankfurt Institut f\"{u}r Informatik, Fachbereich Informatik und Mathematik, Frankfurt, Germany\\
$^{70}$ Korea Institute of Science and Technology Information, Daejeon, Republic of Korea\\
$^{71}$ KTO Karatay University, Konya, Turkey\\
$^{72}$ Laboratoire de Physique des 2 Infinis, Ir\`{e}ne Joliot-Curie, Orsay, France\\
$^{73}$ Laboratoire de Physique Subatomique et de Cosmologie, Universit\'{e} Grenoble-Alpes, CNRS-IN2P3, Grenoble, France\\
$^{74}$ Lawrence Berkeley National Laboratory, Berkeley, California, United States\\
$^{75}$ Lund University Department of Physics, Division of Particle Physics, Lund, Sweden\\
$^{76}$ Nagasaki Institute of Applied Science, Nagasaki, Japan\\
$^{77}$ Nara Women{'}s University (NWU), Nara, Japan\\
$^{78}$ National and Kapodistrian University of Athens, School of Science, Department of Physics , Athens, Greece\\
$^{79}$ National Centre for Nuclear Research, Warsaw, Poland\\
$^{80}$ National Institute of Science Education and Research, Homi Bhabha National Institute, Jatni, India\\
$^{81}$ National Nuclear Research Center, Baku, Azerbaijan\\
$^{82}$ National Research and Innovation Agency - BRIN, Jakarta, Indonesia\\
$^{83}$ Niels Bohr Institute, University of Copenhagen, Copenhagen, Denmark\\
$^{84}$ Nikhef, National institute for subatomic physics, Amsterdam, Netherlands\\
$^{85}$ Nuclear Physics Group, STFC Daresbury Laboratory, Daresbury, United Kingdom\\
$^{86}$ Nuclear Physics Institute of the Czech Academy of Sciences, Husinec-\v{R}e\v{z}, Czech Republic\\
$^{87}$ Oak Ridge National Laboratory, Oak Ridge, Tennessee, United States\\
$^{88}$ Ohio State University, Columbus, Ohio, United States\\
$^{89}$ Physics department, Faculty of science, University of Zagreb, Zagreb, Croatia\\
$^{90}$ Physics Department, Panjab University, Chandigarh, India\\
$^{91}$ Physics Department, University of Jammu, Jammu, India\\
$^{92}$ Physics Program and International Institute for Sustainability with Knotted Chiral Meta Matter (SKCM2), Hiroshima University, Hiroshima, Japan\\
$^{93}$ Physikalisches Institut, Eberhard-Karls-Universit\"{a}t T\"{u}bingen, T\"{u}bingen, Germany\\
$^{94}$ Physikalisches Institut, Ruprecht-Karls-Universit\"{a}t Heidelberg, Heidelberg, Germany\\
$^{95}$ Physik Department, Technische Universit\"{a}t M\"{u}nchen, Munich, Germany\\
$^{96}$ Politecnico di Bari and Sezione INFN, Bari, Italy\\
$^{97}$ Research Division and ExtreMe Matter Institute EMMI, GSI Helmholtzzentrum f\"ur Schwerionenforschung GmbH, Darmstadt, Germany\\
$^{98}$ Saga University, Saga, Japan\\
$^{99}$ Saha Institute of Nuclear Physics, Homi Bhabha National Institute, Kolkata, India\\
$^{100}$ School of Physics and Astronomy, University of Birmingham, Birmingham, United Kingdom\\
$^{101}$ Secci\'{o}n F\'{\i}sica, Departamento de Ciencias, Pontificia Universidad Cat\'{o}lica del Per\'{u}, Lima, Peru\\
$^{102}$ Stefan Meyer Institut f\"{u}r Subatomare Physik (SMI), Vienna, Austria\\
$^{103}$ SUBATECH, IMT Atlantique, Nantes Universit\'{e}, CNRS-IN2P3, Nantes, France\\
$^{104}$ Suranaree University of Technology, Nakhon Ratchasima, Thailand\\
$^{105}$ Technical University of Ko\v{s}ice, Ko\v{s}ice, Slovak Republic\\
$^{106}$ The Henryk Niewodniczanski Institute of Nuclear Physics, Polish Academy of Sciences, Cracow, Poland\\
$^{107}$ The University of Texas at Austin, Austin, Texas, United States\\
$^{108}$ Universidad Aut\'{o}noma de Sinaloa, Culiac\'{a}n, Mexico\\
$^{109}$ Universidade de S\~{a}o Paulo (USP), S\~{a}o Paulo, Brazil\\
$^{110}$ Universidade Estadual de Campinas (UNICAMP), Campinas, Brazil\\
$^{111}$ Universidade Federal do ABC, Santo Andre, Brazil\\
$^{112}$ University of Cape Town, Cape Town, South Africa\\
$^{113}$ University of Houston, Houston, Texas, United States\\
$^{114}$ University of Jyv\"{a}skyl\"{a}, Jyv\"{a}skyl\"{a}, Finland\\
$^{115}$ University of Kansas, Lawrence, Kansas, United States\\
$^{116}$ University of Liverpool, Liverpool, United Kingdom\\
$^{117}$ University of Science and Technology of China, Hefei, China\\
$^{118}$ University of South-Eastern Norway, Kongsberg, Norway\\
$^{119}$ University of Tennessee, Knoxville, Tennessee, United States\\
$^{120}$ University of the Witwatersrand, Johannesburg, South Africa\\
$^{121}$ University of Tokyo, Tokyo, Japan\\
$^{122}$ University of Tsukuba, Tsukuba, Japan\\
$^{123}$ University Politehnica of Bucharest, Bucharest, Romania\\
$^{124}$ Universit\'{e} Clermont Auvergne, CNRS/IN2P3, LPC, Clermont-Ferrand, France\\
$^{125}$ Universit\'{e} de Lyon, CNRS/IN2P3, Institut de Physique des 2 Infinis de Lyon, Lyon, France\\
$^{126}$ Universit\'{e} de Strasbourg, CNRS, IPHC UMR 7178, F-67000 Strasbourg, France, Strasbourg, France\\
$^{127}$ Universit\'{e} Paris-Saclay Centre d'Etudes de Saclay (CEA), IRFU, D\'{e}partment de Physique Nucl\'{e}aire (DPhN), Saclay, France\\
$^{128}$ Universit\`{a} degli Studi di Foggia, Foggia, Italy\\
$^{129}$ Universit\`{a} del Piemonte Orientale, Vercelli, Italy\\
$^{130}$ Universit\`{a} di Brescia, Brescia, Italy\\
$^{131}$ Variable Energy Cyclotron Centre, Homi Bhabha National Institute, Kolkata, India\\
$^{132}$ Warsaw University of Technology, Warsaw, Poland\\
$^{133}$ Wayne State University, Detroit, Michigan, United States\\
$^{134}$ Westf\"{a}lische Wilhelms-Universit\"{a}t M\"{u}nster, Institut f\"{u}r Kernphysik, M\"{u}nster, Germany\\
$^{135}$ Wigner Research Centre for Physics, Budapest, Hungary\\
$^{136}$ Yale University, New Haven, Connecticut, United States\\
$^{137}$ Yonsei University, Seoul, Republic of Korea\\
$^{138}$  Zentrum  f\"{u}r Technologie und Transfer (ZTT), Worms, Germany\\
$^{139}$ Affiliated with an institute covered by a cooperation agreement with CERN\\
$^{140}$ Affiliated with an international laboratory covered by a cooperation agreement with CERN.\\

\end{flushleft} 

\end{document}